\documentclass[11pt,preprint]{emulateapj}
\usepackage{natbib}
\usepackage{xcolor}

\newcommand{\radmm}{~rad m$^{-2}$~}
\newcommand{\absrm}{$|\text{RM}|~$}

\newcommand\pasa{{PASA}}% 

\usepackage{epstopdf}
\usepackage{color}
\usepackage{epsfig}
\usepackage{comment}
\usepackage{amsmath}
\usepackage{graphicx}
\usepackage{threeparttable}
\usepackage{txfonts}

\shorttitle{Jet-gas interactions in the Spiderweb}

\shortauthors{Anderson, Carilli, Tozzi, et al.}
 
\begin{document}

%\title{\rh{[Working title:]} Strong jet-gas interactions in the Spiderweb radio galaxy at redshift 2.16}

\title{The Spiderweb proto-cluster is being magnetized by its central radio jet}

%\correspondingauthor{}
%\email{ccarilli@nrao.edu}

\author{
Craig S. Anderson\altaffilmark{1},
Christopher L. Carilli\altaffilmark{2},
Paolo Tozzi\altaffilmark{3},
G. K. Miley\altaffilmark{4},
S. Borgani\altaffilmark{5, 6, 7, 8},
Tracy Clarke\altaffilmark{9},
L. Di Mascolo\altaffilmark{5},
Ang Liu\altaffilmark{10},
Tony Mroczkowski\altaffilmark{11},
Maurilio Pannella\altaffilmark{5},
L Pentericci\altaffilmark{12},
H.J.A. Rottgering\altaffilmark{4},
A. Saro\altaffilmark{5, 6, 7, 8}
}

\altaffiltext{1}{Jansky Fellow of the National Radio Astronomy Observatory, P. O. Box 0, Socorro, NM 87801, USA, canderso@nrao.edu, ORCID: 0000-0002-6243-7879}

\altaffiltext{2}{National Radio Astronomy Observatory, P. O. Box 0,Socorro, NM 87801, USA, ccarilli@nrao.edu, ORCID: 0000-0001-6647-3861}

\altaffiltext{3}{INAF - Osservatorio Astrofisico di Arcetri, Largo E. Fermi 5, 50125, Firenze, Italy}              

\altaffiltext{4}{Leiden Observatory, Leiden University, P.O.Box 9513, NL-2300 RA, Leiden, The Netherlands}

\altaffiltext{5}{Dipartimento di Fisica dell’ Università di Trieste, Sezione di Astronomia, via Tiepolo 11, I-34131 Trieste, Italy}

\altaffiltext{6}{INAF-Osservatorio Astronomico di Trieste, Via G. B. Tiepolo 11, I-34131 Trieste, Italy}

\altaffiltext{7}{INFN, Sezione di Trieste, Via Valerio 2, I-34127 Trieste TS, Italy}

\altaffiltext{8}{IFPU, Institute for Fundamental Physics of the Universe, via Beirut 2, 34151 Trieste, Italy}  

\altaffiltext{9}{Naval Research Laboratory, Code 7213, 4555 Overlook Ave SW, Washington, DC 20375, USA}         

\altaffiltext{10}{Max Planck Institute for Extraterrestrial Physics, Giessenbachstrasse 1, 85748 Garching, Germany}

\altaffiltext{11}{European Southern Observatory, Karl-Schwarzschild-Str. 2, D-85748 Garching b. Munchen, Germany}

\altaffiltext{12}{INAF - Osservatorio Astronomico di Roma, via di Frascati 33, I-00078 Monte Porzio Catone, Italy }

\begin{abstract}

We present deep broadband radio polarization observations of the Spiderweb radio galaxy (J1140-2629) in a galaxy proto-cluster at $z=2.16$. These yield the most detailed polarimetric maps yet made of a high redshift radio galaxy. The intrinsic polarization angles and Faraday Rotation Measures (RMs) reveal coherent magnetic fields spanning the $\sim60$ kpc length of the jets, while $\sim50$\% fractional polarizations indicate these fields are well-ordered. Source-frame \absrm values of $\sim1,000$ \radmm are typical, and values up to $\sim11,100$ \radmm are observed. The Faraday-rotating gas cannot be well-mixed with the synchrotron-emitting gas, or stronger-than-observed depolarization would occur. Nevertheless, an observed spatial coincidence between a localized \absrm enhancement of $\sim1,100$ \radmm, a bright knot of Ly$\alpha$ emission, and a deviation of the radio jet provide direct evidence for vigorous jet-gas interaction. We detect a large-scale RM gradient totaling $\sim1,000$s \radmm across the width of the jet, suggesting a net clockwise (as viewed from the AGN) toroidal magnetic field component exists at 10s-of-kpc-scales, which we speculate may be associated with the operation of a Poynting-Robertson cosmic battery. We conclude the RMs are mainly generated in a sheath of hot gas around the radio jet, rather than the ambient foreground proto-cluster gas. The estimated magnetic field strength decreases by successive orders-of-magnitude going from the jet hotspots ($\sim90~\muup$G) to the jet sheath ($\sim10~\muup$G) to the ambient intracluster medium ($\sim1~\muup$G). Synthesizing our results, we propose that the Spiderweb radio galaxy is actively magnetizing its surrounding proto-cluster environment, with possible implications for theories of the origin and evolution of cosmic magnetic fields.
\end{abstract}


\section{Introduction}\label{sec:intro}

Supermassive black holes (SMBH) in active galactic nuclei (AGN) interact with the larger cosmos via production of magnetized, radio-emitting synchrotron jets and lobes\footnote{And also through AGN `winds', but this feedback channel is not relevant to the work presented here.} (e.g. \citealp{Gaspari2019,Marsden2020}). This can help drive cosmic ecology and evolution (e.g. \citealp{HC2020}) by controling the flow of SMBH-bound gas on scales ranging from megaparsecs (i.e. in galaxy cluster gaseous halos; \citealt{Gaspari2019} and references therein) down to milliparsecs (i.e. in SMBH accretion disks; e.g. \citealp{EHTC2021,Narayan2021}), by heating galactic and intergalactic gas to limit star-formation and the growth of massive galaxies (e.g. \citealp{Croton2006, Weinberger2017,Weinberger2017b}), or by conversely generating \emph{localized} star formation via compression of gas clouds around the host galaxy (e.g. \citealp{Croft2006,Gaibler2012,Fragile2017,Mukherjee2018}), and by enriching the universe in metals \citep{Reuland2007}, chemical compounds (e.g. \citealp{Russell2017,OSullivan2021}), magnetic fields (e.g. \citealp{FL2001,AG2021}), and cosmic rays (e.g. \citealp{Hardcastle2009,Abdo2010,Eichmann_2018,Vazza_2021}).\\ 

Such interactions may have had their greatest impact on galactic evolution in the approximate redshift range $2<z<4$ (e.g. \citealp{Nesvadba2011,Hatch_2014,Nesvadba2017,Falkendal2019}), the nearside of which sees galactic star formation rates beginning to decline after `cosmic noon' \citep{Schreiber2020}, and a nascent red sequence of quenched galaxies already established \citep{Kriek2008,Brammer2009}. However, the precise roles that radio jets play in driving high$-z$ galactic evolution remain unclear \citep{HC2020}, and directly probing the jet-gas interaction regions at these redshifts remains challenging. Moreover, better-studied low-$z$ radio galaxies are probably poor analogs for their high-$z$ counterparts, since they operate in very different cosmic environments. For example, the mean density of cold gas in high-$z$ galaxies is several times higher than in the present-day universe (e.g. \citealp{Fabian2020}). The energy density of the cosmic microwave background scales as $(1+z)^4$ (meaning that inverse-Compton cooling of radio lobes may scale similarly --- e.g. \citealp{Wu2017,Hodges-Kluck2021,Carilli2022}). The star formation rate is also an order of magnitude greater than present-day levels \citep{MD2014}, galaxy clusters are still in the process of forming (e.g. \citealp{Muldrew2015,Tozzi2022}), and magnetic fields may not yet fully permeate the intergalactic gas \citep{Donnert2018}, which can affect its viscosity, pressure support, and thermal conductivity. Therefore, to better understand how super-massive black holes have shaped the cosmic ecology, it is desirable to observe radio galaxies in nascent galaxy clusters at high redshift, using techniques that illuminate the locus of interaction between the jet and ambient gas, and probe the physics occurring therein.\\

Broadband radio spectropolarimetry represents a singular set of such techniques. Exploiting analysis of Faraday rotation and depolarization, it provides exquisite probes of magnetic field strength and structure around radio-emitting plasma, and by extension, the physical processes operating therein
(e.g. \citealp{CP1962,Burn1966,Conway1974, KS1976,TP1993,Farnsworth2011,OSullivan2013b,JH2015,Gaensler2015,Anderson2015,Anderson2018,Anderson2018b, Anderson2021, OSullivan2012,Anderson2016b,Pasetto2018}). Consider that the linear polarization state of radio emission can be described by a complex vector $\boldsymbol{P}$, related to the Stokes parameters $Q$ and $U$, the polarization angle $\psi$, the fractional polarization $p$ and the total intensity $I$ as

\begin{equation}
\boldsymbol{P} = Q + iU = pIe^{2i\psi}
\label{eq:ComplexPolVec}
\end{equation}

\noindent In traveling from source to observer, linearly polarized radiation will be Faraday rotated by magnetized thermal plasma along the line of sight (LOS) to an observer by an amount equal to

 \begin{equation}
\Delta\psi= \text{RM}\lambda^2
\label{eq:rotation}
\end{equation}

\noindent where $\lambda$ is the observing wavelength, and RM is the Faraday Rotation Measure, which is related to the thermal electron density $n_e$ [cm$^{-3}$] and magnetic field $\boldsymbol{B}$ [$\mu$G] along the LOS as 

\begin{equation}
\text{RM} = 0.812 \int_{z_s}^{0} \frac{n_e(z)B_{||}(z)}{(1+z)^2}\frac{dl}{dz}\text{dz}~~~\text{rad m}^{-2}~,
\label{eq:FaradayDepth}
\end{equation}

\noindent where $z_s$ is the redshift of the radio source, and the comoving path increment per unit redshift, dl/dz, is in parsecs. The RM thus provides a means to detect magnetized thermal plasma, and to help deduce its properties.\\

In terms of high-$z$ proto-cluster/radio galaxy systems to target, there are none more compelling nor better-studied than J1140-2629 (`The Spiderweb Galaxy') at $z = 2.16$ \citep{MD2008, Miley2006}. The radio source is characterized by two powerful jets oriented roughly east-west, extending about 60 kpc from the AGN \citep{Carilli1997,Pentericci1997}. The jets appear to be interacting with gas in the system over a range of scales in multi-faceted ways: The proto-cluster contains a dense agglomeration of galaxies extending to radii $>100$ kpc, whose projected long axis appears to align with that of the radio jets, and whose chain and tadpole morphologies may indicate the radio galaxy has impacted their star formation history (e.g. \citealp{Zirm2005,Miley2006}). Observations of CO emission with the Jansky Very Large Array (VLA) and Australia Telescope Compact Array (ATCA) has revealed a remarkable molecular gas halo, which is also aligned with the radio jet, and which extends to a projected radius of 70 kpc \citep{Emonts2016}. Atacama Large Millimeter Array (ALMA) observations of H$_2$O, CI, and CO emission suggest that the passage of the radio jet can induce condensation of cold molecular gas clouds, intrinsically linking the action of the radio jet with star formation activity in the proto-cluster halo \citep{Gullberg2016}. The entire proto-cluster is enveloped in a spectacular Ly$\alpha$ halo out to a projected radius of $>200$ kpc, which is again broadly aligned with the radio jet \citep{Kurk2002,Miley2006}, and which is strongly displaced by a newly-detected eastern radio lobe \citep{Carilli2022}. An alignment between the radio jet and bright X-ray emission in the system, first described by \citet{Carilli2002}, has been confirmed and delineated in great detail with a new $\sim 700$ ksec Chandra observation \citep{Carilli2022}. The new images show a remarkably close correlation between the radio and the extended X-ray emission from the jet. The data are consistent with the extended X-ray emission arising mostly from inverse-Compton up-scattering of the local photon field by the radio emitting relativistic electrons, with a minor thermal contribution from hot gas.\\

High resolution, broadband, full-polarisation radio observations dis not previously exist, but represent a singular probe of magnetised gas physics in and around the radio jet, and may thereby shed new light on jet-gas interactions in this archetypal protocluster system. We therefore undertook such observations using the Jansky Very Large Array (VLA), which possess sub-arcsecond resolution and span 340 MHz to 36 GHz, as well as a 700 ksec Chandra exposure, as part of a definitive new study of the Spiderweb system \citep{Carilli2022, Tozzi2022}. In this paper, we present an analysis of polarization and Faraday rotation in the Spiderweb system, in order to better understand the properties of the radio jet, the surrounding proto-cluster medium, and interactions between the two, at an epoch where such interactions may peak in their cosmological importance. The paper is laid out as follows: Section \ref{sec:obs} describes our observations and their calibration, and Section \ref{sec:polim} describes our polarimetric imaging and analysis, as well as details of ancillary data sets. Section \ref{sec:results} describes the results derived from our radio observations, and Section \ref{sec:analysis} describes further results obtained via analysis of ancillary data sets. We discuss our findings and conclude in Section \ref{sec:discussion}. We adopt the current Planck cosmology \citep{Planck2020} for our calculations, such that one arcsecond on the sky corresponds to 8.49 kpc at the $z=2.16$ redshift of the source.

\section{Observations and data reduction}\label{sec:obs}

\begin{table*}[!ht]
\caption[Table]{Observing parameters\tnote{a}.}\label{table1}
\centering
\begin{threeparttable}
\small
\begin{tabular}{ll}\toprule
Target & J1140-2629 (`The Spiderweb Galaxy')\\
Project Codes & 19A-024, 20B-448\\
Execution Block IDs & 36425705, 37201837, 37213317, 39133187\\
Dates of observation & 2019-02-28, 2019-09-06, 2019-09-08, 2020-12-06 \\
Field centre (J2000); [$l,b$] & $11^h40^m48.40^s$, $-26^d29^m9.00^s$; [283.872$^\circ$, +33.758$^\circ$]\\
Frequency Bands & X, Ka\\
Frequency Range (X, Ka) & 8--12, 29.2--36.8 GHz\\
Total integration time (X, Ka) & 8, 16 hours\\
Full-band sensitivity (X, Ka) $^*$ & 3, 5 $\muup$Jy beam$^{-1}$ \\
Array configuration(s) (X, [Ka]) & A, [B, BnA-like\textsuperscript{\ref{footnote1}}] \\
Angular resolution (X, Ka; robust $=0.25$) & 0.32"$\times$0.20", 0.31"$\times$0.17" \\
Recorded polarizations & $RR$, $RL$, $LR$, $LL$ \\
$\lambda^2$ range (X, Ka) & 6.25$\times10^{-4}$--1.4$\times10^{-3}$, 6.65$\times10^{-5}$--1.06$\times10^{-4}$ m$^2$\\
Rotation Measure Spread Function width (X; Ka; combined)$^\dagger$ $^\ddagger$ & 4,400; 89,000; 2,600 \radmm \\
Largest recoverable $\phi$-scale$^\dagger$ $^{\ddagger}$ (X; Ka; combined) & 5,000; 47,000; 47,000 \radmm \\
Largest recoverable $|\phi|$ $^\dagger$ $^\ddagger$ $^{\ddagger\ddagger}$ (X; Ka; combined) & 97,000; 1,400,000; 1,400,000 \radmm \\
\hline
\end{tabular}
\begin{tablenotes}
\footnotesize
\item $^*$ Measured per Stokes parameter in multi-frequency synthesis images generated with a Briggs' robust weighting value of +0.25. $^\dagger$ Calculated from equations in Section 6 of \citet{BdB2005}, in the observer frame, to 2 significant figures. Source frame values can be derived by multiplying tabulated values by a factor of $(1+z)^2\approx10$. $^\ddagger$ At greater than 50\% sensitivity. $^{\ddagger\ddagger}$ Calculated at centre frequency of band.
\end{tablenotes}
\label{tab:obsdeets}
\end{threeparttable}
\end{table*}

We observed The Spiderweb Galaxy using the Jansky VLA in multiple bands and array configurations in full polarization for project 19A-024: 2--4 GHz at $1.3"\times 0.6"$ resolution (S-band in A array), 8--12 GHz at $0.4"\times 0.2"$ resolution (X-band in A array), 29.2--36.8 GHz at $0.4"\times 0.2"$ resolution (Ka-band in B array). A total of 8 hours was observed in each band. Additionally, in December 2020, the Jansky VLA was temporarily left in an unscheduled `BnA'-like hybrid configuration due to the COVID-19 pandemic\footnote{\label{footnote1}That is, the northern arm of the array was fully extended. Exact antenna positions are recorded in the measurement sets available from the NRAO data portal: https://data.nrao.edu/portal/}. We sought and obtained 8 hours of director's discretionary time in this configuration in Ka-band, again covering 29.2--36.8 GHz. These supplemental data primarily help improve our Ka-band sensitivity, but also help increase the resolution we would otherwise achieve across the east-west-orientated jet at these southerly latitudes. The final full-band sensitivities of the S-, X-, and Ka-band observations are 5.5, 3, and 5 $\muup$Jy/beam, respectively. The S-band data has been presented in detail by \citet{Carilli2022}; their comparatively low spatial resolution and long wavelength makes them unsuitable for our polarization analysis\footnote{Due to extreme beam depolarization effects, among other reasons}, so they will not be discussed in detail in this work. Further details of our X- and Ka-band observations are listed in Table \ref{tab:obsdeets}.

We used the standard VLA pipeline\footnote{https://science.nrao.edu/facilities/vla/data-processing/pipeline} to flag the data for radio frequency interference (RFI), apply online calibration measurements and flags, and to calibrate the delay, bandpass, flux, and gains. Observations of 3C286 were used to calibrate the bandpass, absolute flux density scale, $RL$-polarization-phase, and absolute polarization angle. J1146-2447 was observed to calibrate the time-dependent gains, and to solve for the on-axis polarization leakage. For the Ka-band observations, J1331+3030 was observed to calibrate the telescope pointing.  

To calibrate the absolute polarization angle and on-axis polarization leakage, we first had to de-apply the pipeline-derived instrument-to-sky-frame Stokes Q and U rotation. We then:

\begin{enumerate}
    \item flagged the cross-hand polarization data manually
    \item used the {\tt CASA} task {\tt setjy} to specify a model for the polarization intensity and angle of 3C286, with Stokes $I$,$Q$,$U$,$V$ set to 1.88, 0.076, 0.233, 0 Jy beam$^{-1}$ (respectively) at a reference frequency of 32 GHz with $\alpha=-0.8$, and noting that $RM\sim0$ \radmm for this source \citep{Perley2013}, as well as the small $\lambda^2$ range spanned by the X- and Ka-bands.
    \item used the above model of 3C286 to solve for the cross-hand delays
    \item applied the above cross-hand delay calibration to our data, then solved for the instrumental D-terms and polarization (simultaneously) every 4 MHz using observations of J1146-2447 repeated over a range of parallactic angle \citep{Sault1996}
    \item applied the above cross-hand delay and leakage calibrations to our data, and then used our model of 3C286 to solve for the instrumental $RL$-polarization-phase every 2 MHz
    \item applied all of the above corrections, which complete the calibration of (on-axis) polarization, to the target data.
\end{enumerate}

To verify the accuracy of the D-term leakage calibration, we applied it to 3C286, whose frequency-dependent polarized fraction is tabulated in \citet{Perley2013}. Their measured average $P/I$ values for the source in the X and Ka bands (respectively) are 12.1\% and 13.3\% ($\pm 0.2$\% systematic), after extrapolation to the present epoch using their Table 5. Our corresponding measurements are 12.4\% and 13.2\%, indicating that our band-averaged leakage calibration accuracy is generally better than $\sim0.3$\% of Stokes $I$.

The off-axis polarization response was not calibrated for these observations. However, the angular extent of the source is small when compared to the full-width-half-maximum of the primary beam at our upper frequency limit ($\sim10$ arcseconds compared to 1.2 arcminutes, respectively), so the source remains squarely in the regime where on-axis polarization effects dominate, and are corrected for. \\

\section{Imaging and analysis}\label{sec:polim}

The pipeline- and polarization-calibrated data were threshold flagged in all polarizations, and self-calibrated in phase (2 rounds) and amplitude (1 round) using {\tt CASA}. We then imaged the data spectropolarimetrically with {\tt WSClean} \citep{Offringa2014}. For all Stokes parameters, we generated channelized images across the full frequency range having $3500\times3500$ pixels, a pixel scale of 0.008 arcseconds, and a \citet{Briggs1995} robust weighting value of 0.25. We imaged and {\sc clean}ed Stokes $I$ MFS maps independently, and then the Stokes $I$, $Q$, and $U$ datacubes using {\tt WSClean}'s `join polarizations' and `squared channel joining' modes, with 100 MHz and 200 MHz channelization in X-band and Ka-band respectively, and using automatic {\sc clean} thresholding and masking (at 1 $\sigma$ and 3$\sigma$ of the full-band noise floor, respectively) based on local noise estimation.

We then smoothed the spectral images to the spatial resolution of our lowest frequency channel --- $0.5\times0.5$ arcseconds --- then re-gridded to a common pixel grid, and concatenated together to form Stokes $I$, $Q$ and $U$ datacubes with dimensions RA, Decl, $\lambda^2$.

We note that the image fluxes were not corrected for the effect of the primary beam (PB), again because the small extent of the source in comparison to the primary beam width means that the magnitude of the correction ($\sim2\%$) is negligible.

We calculated the Faraday Dispersion Spectrum (FDS) in the observer frame over the range $-1\times10^4$ to $+1\times10^4$ \radmm in increments of 250 \radmm using RM synthesis\footnote{\url{https://github.com/brentjens/rm-synthesis}, version 1.0-rc4} \citep{Burn1966,BdB2005} applied to the Stokes $Q$ and $U$ data cubes with equal weighting per image channel. The result is a complex-valued FDS datacube with dimensions RA, Decl., and $\phi$.  

We generated a map of the peak polarized intensity (henceforth ``peak-$P$") and the Faraday-depth-at-peak-polarized-intensity (which for convenience, we henceforth simply refer to as the RM\footnote{These are equivalent for `Faraday simple' sources, where emission within a synthesized beam area can be approximated as coming from a single Faraday depth. See \citet{Anderson2015}, text in Section 3, and Figure 3 caption.}) across the field from the FDS cube using {\tt Miriad}'s \citep{Sault1995} {\sc moment} function, which uses a three-point quadratic fit to the dominant peak of the FDS on a per-pixel basis to derive the amplitude and position of the peak\footnote{https://www.atnf.csiro.au/computing/software/miriad/doc/moment.html}. We verified beforehand that the FDS did not, for the most-part, appear multi-peaked or `Faraday-complex' (see e.g. \citealp{Anderson2015,Anderson2016}) throughout the source. We then masked both the peak-$P$ and peak-$\phi$ maps at a full-band polarized signal-to-noise threshold of $7\sigma$, which is required for reliable RM measurements (e.g. \citealp{Macquart2012}). 

Except where otherwise noted, observer-frame RMs have been ``K-corrected" to the emitting (source) frame through multiplication by a factor of $(1+z)^2$.

\subsection{Ancillary data}\label{sec:ancillary}

\subsubsection{X-rays}\label{sec:xrays}

The diffuse X-ray emission surrounding the Spiderweb Galaxy provides important clues to the properties of the relativistic jets. Such emission includes at least three main components: the wings of the very bright central AGN, Inverse Compton emission in the region overlapping the jets, and thermal emission associated to the proto-cluster gas.  We often refer to this gas as the intracluster medium (ICM), but note that its 
properties may differ from canonical ICM gas in mature clusters at lower redshifts. Dissecting the faint, extended emission in three components, has been possible thanks to the  700 ks Chandra Large Program observation with ACIS-S granted in Cycle 20 (PI P. Tozzi). The details of the ICM, Inverse Compton and AGN X-ray component separation are presented in \citet{Tozzi2022}, while the X-ray emission associated to the radio jets is discussed in detail in \citet{Carilli2022}. 

Briefly though, Tozzi \emph{et al.}'s detection of the proto-ICM is based on a careful characterization of the instrumental and astrophysical background, of the extended wings of the strong unresolved emission of the central AGN, and a spatial identification of the ICM-dominated regions based on the radio data. A thermal X-ray emission component from hot gas (shocked or gravitationally compressed) is thereby detected. Unfortunately, the faintness of this emission, coupled with the uncertainty inherent in the AGN subtraction, means that a detailed radial brightness profile could not be measured for the gas. To first order however, the ICM appears to be roughly isotropic, with a characteristic radius of $\sim 100$ kpc, a volume averaged thermal electron density of $n_e=(1.51\pm0.24\pm0.14)\times 10^{-2}$ cm$^{-3}$ (statistical and systematic uncertainties are quoted, respectively; see \citealp{Tozzi2022}), yielding $M_{ICM}=(1.76\pm 0.30\pm 0.17) \times 10^{12}M_\odot$, and $kT=3.4_{-1.5}^{+4}$ keV, meaning $P_\text{hot}\sim8.2\times10^{-11}$ dyne cm$^{-2}$. These values are used in calculations appearing in Section \ref{sec:analysis}.

\subsubsection{Lyman-$\alpha$}\label{sec:lya}

We use archival data showing Ly$\alpha$ emission from the Spiderweb system (ESO programme 63.O-0477(A), P.I. Miley,  \citealp{Kurk2000}; also see \citealp{Carilli2022} and Nonino et al. 2022, in preparation). Eight exposures totaling four hours of integration time were retrieved from ESO Archive, processed to remove instrumental signatures, then stacked using {\tt swarp} \citep{Bertin2002}. The narrow band filter has a central wavelength of 381.4 nm and a full-width-at-half-maximum (FWHM) of 6.5 nm. The diffuse emission in this filter is dominated by the rest frame Ly$\alpha$ emission from J1140-2629. Some underlying continuum emission also remains, but is negligible for the purposes of this work.

\section{Results}\label{sec:results}

\subsection{Total intensity}\label{sec:results_totintens}

\begin{figure} % Gen with spiderweb_X-band_hueint_alphablend_Lya.py
\centering
\includegraphics[width=0.5\textwidth]{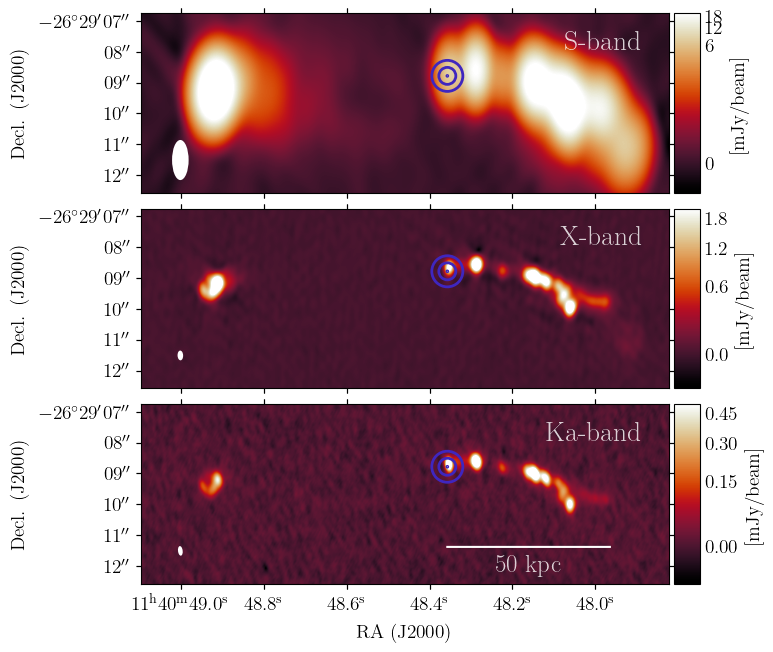}
\caption{Multi-frequency synthesis maps of the Spiderweb radio jet and lobes at S-band (row $a$), X-band (row $b$), and Ka-band (row $c$). The full source is shown for scale and perspective. All maps are shown on the same coordinate scale. The X- and Ka-band images are presented with square-root-stretched color scales. The S-band image is presented with a logarithmic color scale to emphasize detail over a larger dynamic range. For all images, the sensitivity is approximately 5 $\muup$Jy/beam. The synthesized beam areas are indicated at the bottom-left in the right-hand column of plots. The position of the X-ray AGN core \citep{Tozzi2022} is indicated with blue concentric circles. The angular scale is identical between images; the white scale-bar indicates a projected span of 50 kpc at $z=2.16$ for the adopted cosmology.}
\label{fig:s-x-ka-total-int}
\end{figure}

Figure \ref{fig:s-x-ka-total-int} shows S-, X-, and Ka-band multi-frequency synthesis (MFS) total intensity radio maps of the system. These are described in detail by \citet{Carilli2022}. For this work, the features of interest in the total radio intensity maps are (a) the radio core, which is the eastern-most radio component visible in the western jet, and which is so-identified due to its spatial coincidence with the bright X-ray-emitting AGN (see Figure 2 of \citealt{Carilli2002}), (b) a deflection and subsequent bifurcation of the western jet, which is co-located with bright knots of both radio emission and Ly$\alpha$ emission (also described by \cite{Kurk2002}), and see Section \ref{sec:RMenhancement}), and (c) the relatively complex morphology of the eastern hotspot complex, which is resolved in the X and Ka bands. We note that the latter is embedded in a newly-detected eastern lobe described by \citet{Carilli2022}, which is faintly visible in the S-band map.

\subsection{Polarization}\label{sec:results_pol}

\begin{figure} % Gen with spiderweb_X-band_hueint_alphablend_Lya.py
\centering
\includegraphics[width=0.5\textwidth]{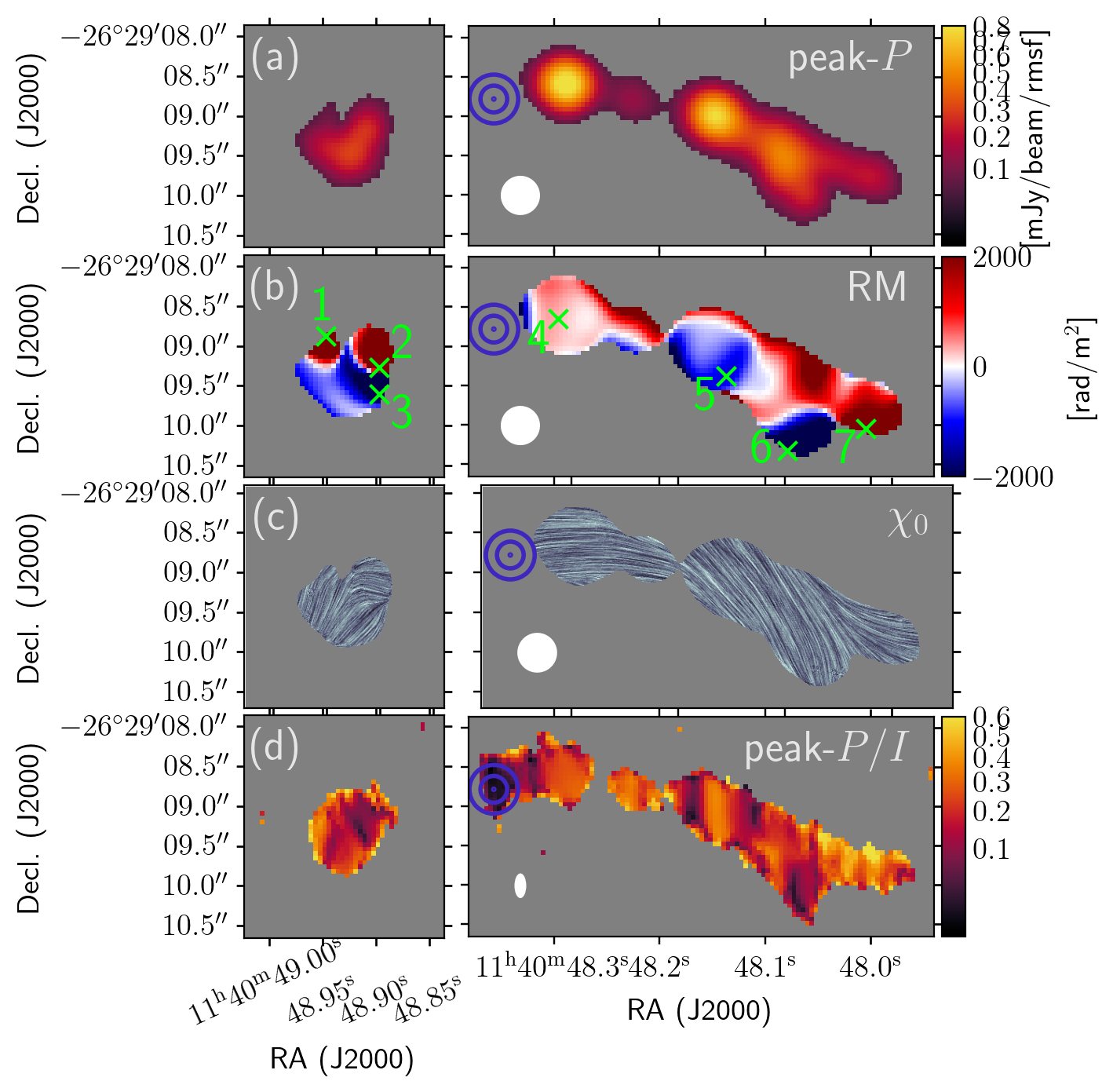}
\caption{Maps of peak-$P$ (X- plus Ka-band), source-frame RM (X- plus Ka-band; note that values saturate at $\text{RM}=\pm2000$ \radmm to better reveal low \absrm structure throughout the jet), intrinsic sky-projected magnetic field orientation (Ka-band only), and fractional polarization (Ka-band only) across the Spiderweb radio jet (rows a--d respectively). The eastern and western jet components (see Figure \ref{fig:s-x-ka-total-int}) are shown in the left-hand and right-hand columns respectively. Coordinates are shared between axes in the same way as for Figure \ref{fig:s-x-ka-total-int}. The effective synthesized beam areas (see Section \ref{sec:polim}) are shown in the lower-left corner of the right-hand column of panels in white. The location of the X-ray AGN core \citep{Tozzi2022} is indicated with blue concentric circles. The maps have been masked at full-band polarized signal-to-noise levels of $7\sigma$ (for their respective frequency band coverage). The map of fractional polarization has been masked at a full-Ka-band signal-to-noise of $10\sigma$ in the total intensity. The green 'x' markers on the RM map mark the positions at which RMs were extracted for values presented in Table \ref{tab:extremerms}. The uncertainties in RM (row $b$), intrinsic magnetic field orientation (row $c$), and fractional polarization (row $d$) typically lie in the range 100--400 \radmm, 0--5 degrees, and 0--0.03, respectively.}
\label{fig:p_rm_ang}
\end{figure}

\begin{figure*} % Gen with spiderweb_X-band_hueint_alphablend_Lya.py
\centering
\includegraphics[width=0.99\textwidth]{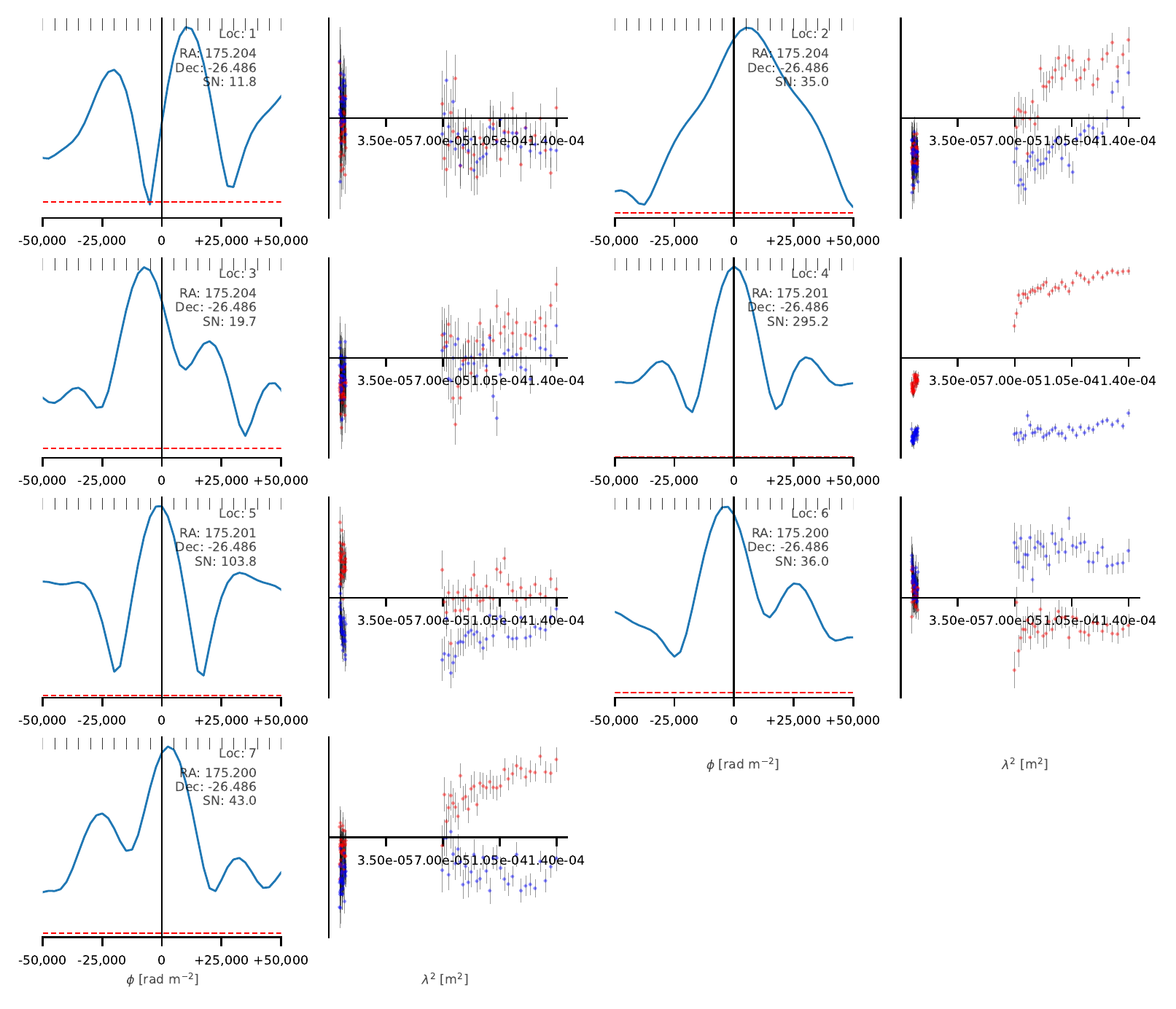}
\caption{The calculated dirty (i.e. no {\tt rmclean} performed; \citealp{Heald2009}) FDS (first \& third columns) and corresponding Stokes $Q$ (red) and $U$ (blue) spectra (second \& fourth columns) for the seven high-\absrm locations labeled in Figure \ref{fig:p_rm_ang} panel (b), and listed in Table \ref{tab:extremerms}. For the FDS plots, the horizontal axes range from -50,000 to +50,000 rad m$^{-2}$ in the source frame. The gray minor tick marks at the top of each panel indicate increments of 5,000 \radmm. For the $QU$ plots, the horizontal axes span 0--$1.4\times10^{-4}$ m$^2$, again in the source frame. The vertical axes limits are all scaled to the maximum amplitude of the data points in individual plots. The location number (corresponding to those shown in Figure \ref{fig:p_rm_ang} panel (b) and Table \ref{tab:extremerms}), the right ascension, declination, and band-averaged polarized signal-to-noise ratio (SN) are all written in the respective FDS plots. The error bars on the ($Q$,$U$) data points indicate the standard deviation measured per image channel from the Stokes ($Q$,$U$) datacubes in a small region adjacent to each source. Note that because the FDS have not been deconvolved with {\tt rmclean}, the emission-free regions of the FDS cannot be used as a guide to the underlying noise level. Note that the broad FDS peak seen at `location 2' may be a rare example (see Section \ref{sec:polim}) where emission comes from a range of Faraday depths (i.e. shows `Faraday complexity'). This is unsurprising, given its location near RM `interfaces' in the eastern hotspot complex.}
\label{fig:specpoldata}
\end{figure*}

Figure \ref{fig:p_rm_ang} presents maps of peak-$P$ (panel row $a$), the source-frame RM (row $b$), intrinsic polarization angle (row $c$), and fractional polarization (row $d$) across the system, derived from a mixture of X- and Ka-band spectral data (specified in caption). Figure \ref{fig:specpoldata} presents raw spectropolarimetric data for selected locations in the system, further described in Section \ref{sec:extremeRMs}.

The appearance of the system is similar in both polarized and total intensity. An obvious difference is that the radio core is not detected in polarization. The upper limit on the core's Ka-band fractional polarization is $\sim0.5\%$, but as we move westwards down the jet, the fractional polarization rapidly increases to $\sim30\%$ or more, after which the jet shows a complex mix of high and low fractional polarization values ranging up to 50\%. The (X- plus Ka-band) RMs show similarly complex structure on a variety of scales throughout the eastern hotspot complex and western jet, down to scales we describe as `interfaces' (e.g. see \citealp{Anderson2018}), where the RM shows large changes in both magnitude and sign over scales smaller than the synthesized beamwidth. Conversely, the sky-projected magnetic field changes only slowly, and is broadly aligned with the jet axis over the entire length of the system. We investigate these basic features of the system in more detail below.

\subsubsection{The magnetic structure of a high redshift radio galaxy}\label{sec:detailedmapB}

%, and $281^\circ$ versus $258^\circ$ in the western

Our observations provide the most detailed ever maps of the magnetic field structure in a high redshift source. The emission-weighted, sky-projected magnetic field orientation (Figure \ref{fig:p_rm_ang} panel c) is broadly aligned with the jet axis over the entire length of the system ($\sim70$ kpc) where polarized emission is detected. At the beginning of the western jet (between locations $4$ and $5$ in Figure \ref{fig:p_rm_ang} panel $b$), the position angles of the jet and magnetic field are $261^\circ$ versus $260^\circ$ (respectively, on average, increasing eastwards from north, and with an uncertainty of $\sim5$ degrees on the quoted measurements). Post-bifurcation, the jet consists of both a ``southern" and ``western" branch. In the southern branch (roughly between positions 5 and 6 in Figure \ref{fig:p_rm_ang} panel $b$) the measured orientation angles are $227^\circ$ versus $210^\circ$, and likewise $281^\circ$ vs. $258^\circ$ in the western jet extension. The eastern hotspot lies at a position angle of $95^\circ$ from the nucleus. While the projected magnetic field orientation is relatively more structured in this location, its mean position angle is $120^\circ$. Thus, for both the eastern and western jets, the magnetic field orientation lies predominantly along the jet, but with a minor cross-jet component in some locations (of between 0 and 30 degrees).

The southern jet branch appears brighter and broader in total intensity than the western jet branch, but terminates at a bright component or hotspot after less than an arcsecond. The latter is fainter and narrower, but appears to continue for two or more arcseconds, and terminates in a faint lobe that is visible in our maps at S-band, but only faintly so at X-band (see Figure \ref{fig:s-x-ka-total-int}). The western branch is characterized by positive RM and higher fractional polarization compared to the southern branch (Figure \ref{fig:p_rm_ang}, panels b and d respectively), the latter of which suggesting that the magnetic field is more ordered here, and perhaps stronger. The emission-weighted magnetic orientation vectors appear to change smoothly into the brighter post-bifurcation southern branch from the pre-bifurcation main branch, but show an abrupt change moving from the former into the western branch. This is likely because the emission-weighted magnetic orientation vectors are dominated by the brighter southern branch, until the two branches separate completely on the plane of the sky.

Assuming that Faraday depolarization is negligible in the Ka-band, the observed fractional polarization of the jet carries information about the intrinsic degree of ordering of the plane-of-sky-projected magnetic field, with $p_\text{obs}$ related to the degree of field ordering as $p_\text{obs}=p_\text{max}(1+B_\text{r}^2/B_\text{o}^2)^{-1}$ \citep{Burn1966}, where $p_\text{max}\sim0.75$ is the theoretical maximum fractional polarization generated by synchrotron radiation, $B_\text{r}$ is the strength of the random component of the decomposed magnetic vector field, and $B_\text{o}$ is likewise the strength of the ordered or regular component. The observed fractional polarization at Ka-band typically ranges between 25--40\% in the western jet, and 10--25\% in the eastern hotspot complex, so that the ratio $B_\text{r}/B_\text{o}$ ranges from 0.9--1.4 in the western jet, and from 1.4--2.5 in the eastern hotspot complex. Generally then, the magnetic field structure of the jet is quite well-ordered, with a substantial poloidal (along-jet) component. There are two classes of models that purport to explain the presence of this poloidal component -- i.e. those that invoke (i) helical fields and a roughly cylindrical jet \citep{Lyutikov2005}, and (ii) velocity shear (e.g. \citealp{Laing1980, HAA1989a, HAA1989b,HAA1991,Attridge1999,Kharb2005}). Our analysis in Section \ref{sec:RMgradients} suggests that in the present case, scenario (i) applies in a sheath-like region around the jet, but is agnostic about scenario (ii).

%For frac pol analysis above, see kvis k-MFS-PonI-immath_derived.fits in /Users/canderso/Data/JVLA/spiderweb/ka-all-new-specpol-qufitting-rob0p6_16chan/images/

%\rh{ (from https://academic.oup.com/mnras/article/415/3/2081/1044603): "Two classes of models explain these data: (i) models assuming a helical field and a cylindrical jet (Lyutikov et al. 2005) and (ii) shock and velocity shear models which assume that shock compression along the jet axis produces EVPAs parallel to the jet axis and that velocity shear explains EVPAs perpendicular to the jet axis (Laing 1980; Hughes, Aller \& Aller 1989a,b, 1991; Attridge, Roberts \& Wardle 1999; Kharb, Shastri \& Gabuzda 2005)."}

\subsubsection{An updated search for Extreme Faraday rotation measures}\label{sec:extremeRMs}

A map of the intrinsic (source-frame) RMs calculated from the combined X- and Ka-band data is shown in Figure \ref{fig:p_rm_ang} (row $b$). Selected locations are marked with green crosses; corresponding measurements are presented in Table \ref{table2}, and FDS and raw $QU$ spectra in Figure \ref{fig:specpoldata}. 

We detect intrinsic \absrm values consistent with those reported by \citet{Carilli1998} for the eastern hotspot complex, but also detect similarly large \absrm values throughout several fainter regions of the jet. In the north-most part of the eastern hotspot complex, we measure an \absrm value of $11,100\pm400$ \radmm, which is roughly double the previously largest reported value. It is noteworthy that the most extreme \absrm values in the system tend not to occur in the central regions of the proto-cluster, but are found (a) towards the eastern and western terminal points of the radio jet (see Table \ref{table2}), and (b) towards the edges of the jets in general. The latter is relevant to results in Section \ref{sec:RMgradients}.

We also detect interfaces in RM, where the value changes dramatically over spatial scales smaller than the synthesized beam width --- from $+5,800\pm180$ \radmm to $-4,400\pm360$ \radmm in the most extreme case visible in the eastern hotspot complex. The RM values often cross through zero over these interfaces, which implies an associated change in the line-of-sight-projected magnetic field direction. The interfaces may also be associated with some of the most heavily depolarized regions visible in the lobes at Ka-band fractional polarization map (Figure \ref{fig:p_rm_ang}, row $d$). Given the frequency, the depolarization is unlikely to be due to Faraday effects per se, but rather geometric depolarization arising from associated complex magnetic field structures in the emitting region. Similar structures have been described in Fornax A, where they were found to be associated with complex internal lobe structure and entrained thermal gas \citep{Anderson2018}.\\

Because extreme RM values often coexist with extreme RM dispersion, the most extreme \absrm values are often best sought by restricting analysis to the highest available frequency bands. Otherwise, the signal can be overlooked due to strong depolarization, or overwhelmed by strong polarized emission emerging from lower-RM regions falling along the same sight lines. We therefore conducted a search for even higher \absrm values than reported above using the Ka-band data only. This analysis did not reveal RM values significantly different from those shown in Figure \ref{fig:p_rm_ang} (row $b$), though we note that the Ka-band-only source-frame RM uncertainties are typically thousands of \radmm for the sensitivity attained in these observations. 

%At such high frequencies, source frame RM dispersions of $\sim1.5\times10^4$ \radmm are required to produce even moderate depolarization at the upper end of the band, making complete depolarization an unlikely problem. 

\begin{table*}[!ht]
\caption[Table]{Source-Frame Faraday Rotation Measures for Selected Locations in the Spiderweb Radio Jet\tnote{a}.}\label{table2}
\centering
\begin{threeparttable}
\small
\begin{tabular}{ccrc}\toprule
Location index (Fig. \ref{fig:p_rm_ang}, panel \emph{b})$^a$ & Coordinates & RM (source frame) & Projected Distance from AGN\\
  & (J2000) & [\radmm] & [kpc]\\
\hline
1 & $11^h40^m48.945^s$, $-26^d29^m08.86^s$ & $+11,100\pm400$ & 66.2\\
2 & $11^h40^m48.895^s$, $-26^d29^m09.26^s$ & $+5,800\pm180$ & 60.5\\
3 & $11^h40^m48.895^s$, $-26^d29^m09.59^s$ & $-4,400\pm360$ & 60.7\\
\hline
4 & $11^h40^m48.294^s$, $-26^d29^m08.64^s$ & $+270\pm30$ & 9.0\\
5 & $11^h40^m48.135^s, -26^d29^m09.37^s$ & $-1,200\pm100$ & 27.3\\
6 & $11^h40^m48.077^s$, $-26^d29^m10.32^s$ & $-4,400\pm390$ & 35.8\\
7 & $11^h40^m48.002^s$, $-26^d29^m10.04^s$ & $+3,500\pm270$ & 43.5\\
\hline
\end{tabular}
\begin{tablenotes}
\item[a] Note that \absrm values shown in Fig. \ref{fig:p_rm_ang}, panel \emph{b} saturate at 2000 \radmm as per the figure caption
\end{tablenotes}
\label{tab:extremerms}
\end{threeparttable}
\end{table*}

%\subsection{A Rotation Measure enhancement co-spatial with a bright Lyman-$\alpha$ knot, radio jet bifurcation and deviation}\label{sec:RMenhancement}

\subsubsection{A co-spatial Lyman-$\alpha$ knot, jet bifurcation, and Rotation Measure enhancement}\label{sec:RMenhancement}

\begin{figure} % Gen with spiderweb_X-band_hueint_alphablend_Lya.py
\centering
\includegraphics[width=0.5\textwidth]{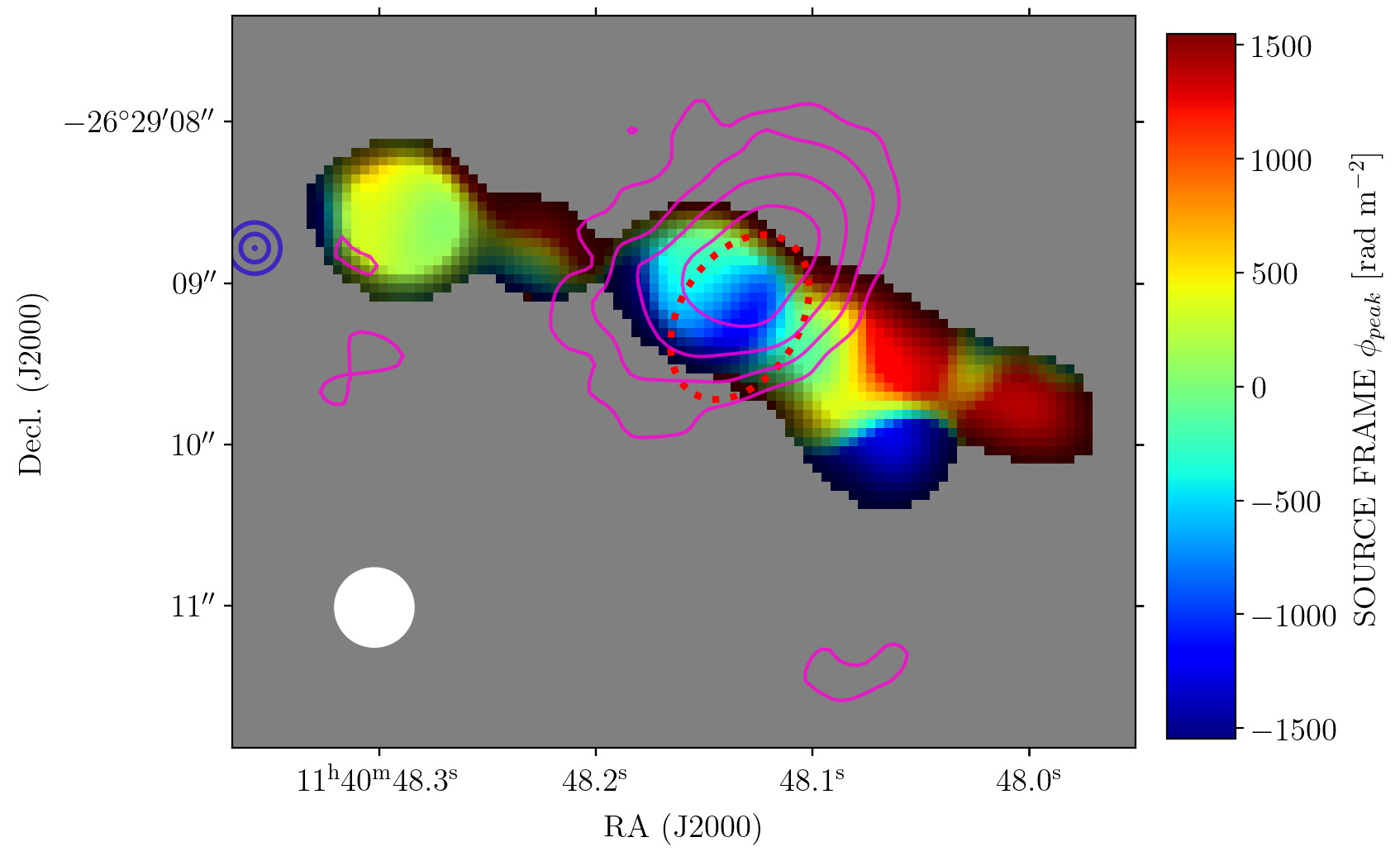}
\caption{Colorscale: Hue-intensity map of the western radio jet. The hue channel traces source-frame RM, while the intensity channel traces peak-$P$. The color map has been selected to highlight local differences in RM. Masking is the same as Figure \ref{fig:p_rm_ang}. A flat Galactic foreground contribution of $-13$ \radmm has been subtracted from $\phi_\text{peak}$ \citep{HE2020}. Contours: Lyman-$\alpha$ emission. Levels start at 35 arbitrary units, and increase by a factor of $\sqrt{2}$. The red dotted line indicates the location of the RM enhancement discussed in the main text. The position of the X-ray AGN core \citep{Tozzi2022} is indicated with blue concentric circles. The white filled circle indicates the effective synthesized beam size.}
\label{fig:fd_lya_olay}
\end{figure}

Early studies of the Spiderweb system (e.g. \citealp{Pentericci1997,Kurk2002}) highlighted that the western radio jet changes course and bifurcates at the location of a bright knot of Ly$\alpha$ emission. Figure \ref{fig:fd_lya_olay} shows that the RM is also enhanced at this location, jumping from $|\text{RM}|=200\pm40$ \radmm to $|\text{RM}|=1200\pm50$ \radmm at the position of the Ly$\alpha$ knot (moving WSW down the jet), and then immediately dropping back to $|\text{RM}|=0\pm80$ \radmm --- corresponding to an $\sim1100$ \radmm enhancement (relative to the average of the \absrm values measured immediately upstream and downstream of the knot) spanning $\sim1.5$" (12.75 kpc). Whilst larger \absrm enhancements occur in other locations in the jet, this particular enhancement distinguishes itself by virtue of (a) being spatially coincident with the bright knots of both Ly$\alpha$ and polarized radio emission, as well as the jet bifurcation and deviation as previously noted, (b) being immediately bracketed by two regions with much lower absolute RM values, and (c) the direction of steepest \absrm gradient to and from this location being oriented along the jet axis, rather than perpendicular to the jet axis, as it generally is for other notable RM enhancements (see Section \ref{sec:RMgradients}). We conclude that the jet impinges on the Ly$\alpha$ emission region, compresses the gas and magnetic fields in this location, and thereafter deviates from its course and bifurcates. n Section \ref{sec:Lya_ancillary}, we use ancillary data (Section \ref{sec:ancillary}) to constrain the nature of the Faraday-active medium associated with this Ly$\alpha$ knot, and calculate the conditions in the gas.

\subsubsection{Transverse Faraday rotation measure gradients}\label{sec:RMgradients}

\begin{figure} %Gen with something like spiderweb_multinest_multicore_fitter.py
\centering
\includegraphics[width=0.45\textwidth]{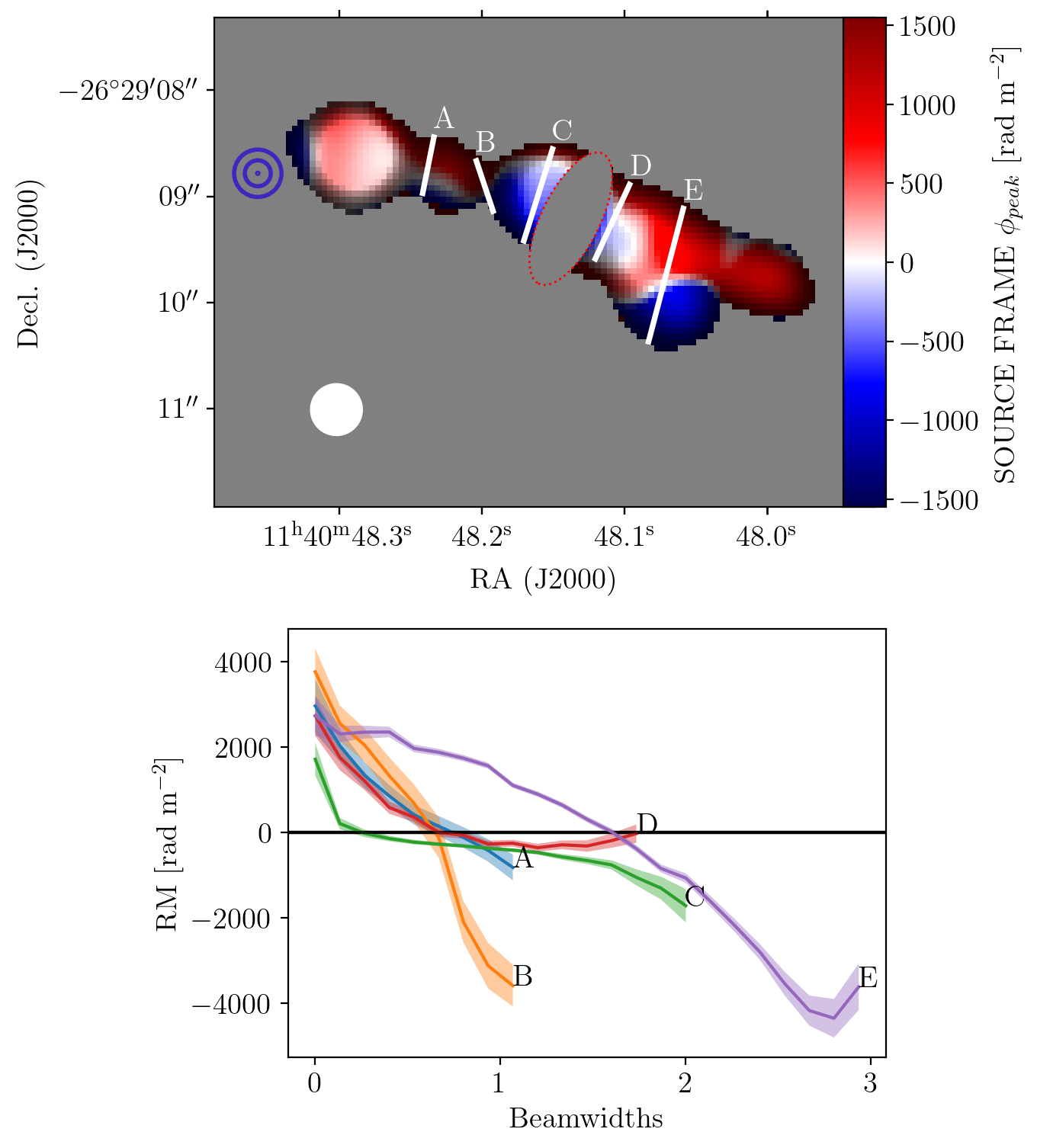}
\caption{\emph{Top panel}: Hue (source-frame RM)-intensity (peak-$P$) map of western radio jet, as for Figure \ref{fig:fd_lya_olay}. A divergent color map has been selected to highlight areas where RMs are zero \radmm (white), positive (red) and negative (blue). Masking is the same as Figure \ref{fig:p_rm_ang}, with the addition of a red-dotted ellipse, which also masks the peak-$P$ enhancement attributed to local interaction with the Ly$\alpha$ knot (described in Section \ref{sec:RMenhancement}). This emphasizes the surrounding global RM structure of the jet. Five cross sections running across the axis of the western jet, and spaced roughly equidistantly along the jet, are indicated by white lines and lettered designations. The location of the X-ray AGN core \citep{Tozzi2022} is indicated with blue concentric circles. The white filled circle indicates the effective synthesized beam size. \emph{Bottom panel}: Values of source-frame RM (colored lines) and their uncertainty (shaded regions), plotted as a function of angular distance in units of synthesized beamwidths, extracted from north to south along the cross-sections indicated by the corresponding letter designations in the top panel. }
\label{fig:rm_gradients_west_jet}
\end{figure}

Figure \ref{fig:rm_gradients_west_jet} (top panel) shows the same hue-intensity map (tracing source-frame RM and peak-$P$, respectively) as Figure \ref{fig:fd_lya_olay}, but presented with a diverging binary color map, and the \absrm enhancement/Ly$\alpha$ knot region masked, to emphasize structure that we will now discuss. It evident that regions of the jet northward of main east-west jet axis show predominantly positive RMs, whereas the converse is true southward of the jet axis. Moreover, larger \absrm values in the system tend to be found towards the edges of the jet (Section \ref{sec:extremeRMs}). To illustrate these effects, we extracted the RMs along the cross sections labeled A--E in the top panel of Figure \ref{fig:rm_gradients_west_jet}, and plot the results in the bottom panel of the same figure. In each case, the RM along the cross sections show large, monotonic decreases in RM, which typically cross through zero. While the cross sections only extend over between 1 and 3 total intensity synthesized beamwidths, we are confident that the RM gradients are genuine because:

\begin{enumerate}
\item polarization is a vector quantity, meaning that structures smaller than the FWHM of the total intensity synthesized beam can and will be revealed in RM maps (in fact, the same is true of total intensity, given sufficient signal strength --- e.g. \citealp{Dabbech2018})
\item Monte Carlo simulations in the very long baseline interferometry (VLBI) literature show that RM gradients can be reliably \emph{detected} over even a small fraction of a beamwidth --- it is the statistical significance and monotonicity of the change in RM that are most important in this key respect \citep{Murphy2012,Hovatta2012,Mahmud2013}.
\item the RM gradients are not confined to a single poorly-resolved knot of emission (as has been the case for some of the RM gradients claimed to exist in the VLBI literature), but rather exist along almost the entire length of the western jet 
\item the extracted RM cross-sections plotted in Figure \ref{fig:rm_gradients_west_jet} are recovered with similar maximum magnitude ($\sim2000-4000$ \radmm), RM span ($\sim4000-8000$ \radmm) and orientation (increasing most rapidly from south to north across the jet axis), and in three cases span two or more synthesized beamwidths. None of this is expected if the gradient structure is in fact produced by random unresolved noise (e.g. \citealp{Hovatta2012}). 
\end{enumerate}

\noindent Finally, we note that we performed experiments where we intentionally offset the X-band and Ka-band polarization maps by several pixels in both RA and Decl, to assess whether self-calibration-induced astrometric offsets could spuriously cause the observed RM gradients. The RM gradients remained unchanged.\\

Thus we claim to have robustly detected these RM gradients. Their presence represents additional evidence to constrain the nature of the Faraday rotation observed throughout the system. The RM reversal across the gradients implies a change in the  line-of-sight-projected magnetic field direction. The effect is observed along at least $\sim20$ kpc of the western jet, discounting the local change in RM attributed to the Ly$\alpha$ knot in Section \ref{sec:RMenhancement}. It follows that the RMs are generated in the immediate vicinity of the jet --- it seems implausible that the unassociated ICM in the foreground `knows' about the axis of the radio jet, and is configured to generate zero RM precisely along this locus of points, with predominantly positive and negative RM above and below this line, respectively. On the other hand, the synchrotron-emitting particles in the jet are probably not well-mixed with the Faraday-rotating medium, since relevant depolarization models (like the `Burn slab' or internal Faraday dispersion; see e.g. \citealt{OSullivan2012}) generally predict polarized fractions of less than 10\% at X-band for measured \absrm values greater than $\sim1000$ \radmm. This is lower than what we observe throughout much of the jet. We conclude that the Faraday rotation probably originates in a magnetized cocoon around the jet. 

\begin{figure} %Gen with something like spiderweb_multinest_multicore_fitter.py
\centering
\includegraphics[width=0.45\textwidth]{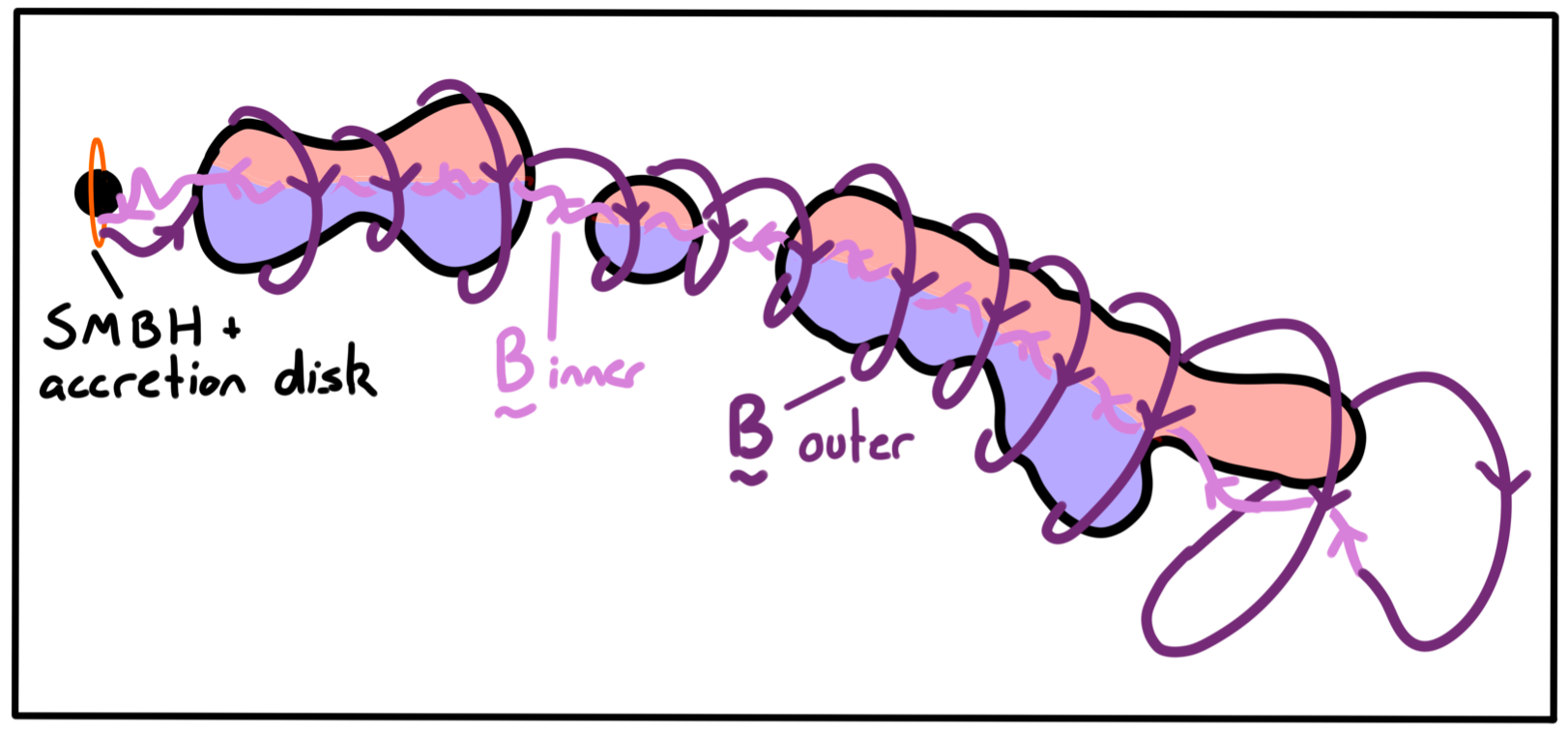}
\caption{Toy model of the western jet (black outline), indicating how an RM gradient crossing from negative RMs (blue shading) to positive RMs (red shading; similar to Figure \ref{fig:rm_gradients_west_jet}) across the jet axis could arise from a toroidal magnetic field component which is clockwise when viewed from the central SMBH and accretion disk (labelled). The picture is of a nested helix configuration, with inner and outer magnetic field windings (labelled), as discussed in Section \ref{sec:discussion}).}
\label{fig:toy_model}
\end{figure}

The VLBI literature argues that significant and monotonic RM gradients typically imply the presence of helical or toroidal magnetic fields that are directly coupled with the radio jets (e.g. \citealp{Mahmud2013, Gabuzda2018,Gabuzda2018b}). While this type of analysis has most often been performed on parsec scale jet emission, there is emerging evidence that RM gradients are also observed on kiloparsec scales in some sources \citep{Christodoulou2016,Knuettel2017}. The sense of kpc-scale RM gradients may be an important probe of accretion disk physics, as we discuss in Section \ref{sec:discussion}. We therefore note that in the present case, the observed RM gradient increases from south to north in the western jet, indicating that the toroidal component of the magnetic fields have a clockwise orientation when viewed along the jet from the central AGN (Equation \ref{eq:FaradayDepth}; also see Figure \ref{fig:toy_model}). 

\section{Analysis incorporating ancillary results}\label{sec:analysis}

\subsection{Magnetic fields in the Ly$\alpha$ knot}\label{sec:Lya_ancillary}

In Section \ref{sec:RMenhancement}, we reported on the discovery of a $\sim1100$ \radmm Faraday depth enhancement in the radio jet associated with a Ly$\alpha$ knot. By considering ancillary data and results previously reported from such (Section \ref{sec:ancillary}), we now consider the nature of the Faraday-active medium. Assuming a Ly$\alpha$ volume filling factor of $f\sim1\times10^{-5}$ \citep{McCarthy1993,Pentericci99} implies that the length filling factor of this gas phase is $l=1\times10^{-5/3}=2.2\times10^{-2}$, while the areal filling factor is only $a=1\times10^{-10/3}=4.6\times10^{-4}$. Therefore, fewer than 1 in 2000 radio photons from the jet will pass through the Ly$\alpha$-emitting gas itself, ruling it out as an important source of the observed Faraday rotation. Nevertheless, the warm and dense Ly$\alpha$ cloudlets must be confined by a pervasive external plasma in approximate pressure equilibrium. \citet{Pentericci1997} estimate $n_{e,Ly\alpha}\sim40$ cm$^{-3}$ \footnote{Based on the measured flux and extent of the Ly$\alpha$ emitting gas, and assuming this gas phase predominantly arises via photoionization from massive stars. This value may only be accurate to a factor of $\sim$a few, but is within the 10--100 cm$^{-3}$ range derived for similar systems at $z\gtrsim2$} (e.g. \citealp{McCarthy1990,Chambers1990,vO1995,VM2003,Falkendal2021}), so assuming $T_{Ly\alpha}\sim10^4$ K, the Ly$\alpha$-emitting gas will then have a pressure $P_{Ly\alpha} = n_ekT = 5.5\times10^{-11}$ dyne cm$^{-2}$. The most plausible confining agent is a hot X-ray-emitting cluster gas, which has now been detected by \citet{Tozzi2022}. They derive $P_\text{hot}\sim8.2\times10^{-11}$ dyne cm$^{-2}$ (Section \ref{sec:xrays}), which is within 50\% of the $P_{Ly\alpha}$ value estimated above --- consistent with a state of pressure equilibrium within the substantial uncertainties, and a viable candidate for the Faraday-active material. Exploiting Tozzi \emph{et al.}'s estimate that $n_{e,hot}\approx0.015$ cm$^{-3}$ (Section \ref{sec:icm}), and assuming that (i) the hot gas has unit filling factor throughout the Ly$\alpha$-emitting knot, (ii) the magnetic field is uniform over this region, and (iii) minimal Faraday rotation occurs in the foreground (justifiable given that the RM enhancement and Ly$\alpha$ knot are similar in size ($\sim10$ kpc); also see Section \ref{sec:icm}), we use Equation \ref{eq:FaradayDepth} to estimate that $B_\text{hot}\sim9~\muup G$ in the vicinity of the interaction region. This is comparable to results for radio galaxies in dense clusters at low redshift \citep{CT2002}. The implied magnetic pressure is $P_B=B^2/8\pi=3.2\times10^{-12}$ dyne cm$^{-2}$, which is only $\sim5\%$ of the thermal pressure in both the Ly$\alpha$-emitting gas and its confining medium. The assumptions inherent in such calculations limits accuracy to a factor of $\sim$several, and these uncertainties are effectively irreducible for the types of measurements under consideration (e.g. \citealp{Johnson2020}), so the various pressure values are in fact broadly comparable. \\

%NOTE TO SELF: Numbers derived above are changed from the original draft which calculated n_e=0.017 cm^-3, rather than simply using the Tozzi number of 0.012 cm^-3. Flow on effects have been scaled on this basis, and so will not match the original python script.  

It is interesting to speculate whether the jet-gas interaction causes a velocity disturbance that is measurable with spectroscopy, and whether such measurements could provide a more detailed accounting of energetics and momentum transfer along the western jet. Integral Field Unit (IFU) spectroscopic data would be ideal for this purpose. However, none yet exist that cover the red-shifted Ly$\alpha$ line in this system. IFU observations of other emission lines do exist \citep{Kuiper2011}, but the relationship between the different gas phases is unclear, and a detailed consideration of such is not warranted here. Nevertheless, Figure 2 of \citet{Kurk2002} show hints of an associated velocity disturbance. The spatial coverage of their long-slit spectra does not include the main part of the Ly$\alpha$ knot described above (and labeled component `d' in their Figures 1 \& 2), but does run adjacent to it and covers part of its edge. In this location, their Figure 2 shows that the total velocity dispersion ranges over $\sim3000$ km s$^{-1}$, while the brightest emission is red-shifted by $\sim1000$ km s$^{-1}$ from $z=2.16$, and by $\sim2000$ km s$^{-1}$ from the nuclear Ly$\alpha$ emission (see also \citealp{Kuiper2011}). These values are broadly comparable to the $\sim1600$ km s$^{-1}$ eastern hotspot advance speed estimated by \citet{Carilli2022}, which suggests that the jet-gas interaction is similarly vigorous in both locations.

\subsection{A magnetized ICM?}\label{sec:icm}

Cluster ICMs are generally observed to be magnetized in the local universe (e.g. \citealp{Anderson2021}; \citealp{Heald2020} and references therein), which is often invoked to explain the RM structure of radio jets embedded therein (e.g. \citealp{Dreher87,Murgia2004,VE2005,Bonafede2010}, but see \citealt{Anderson2018}). But can we assume this equally applies to the Spiderweb system and its (proto-)ICM at $z=2.16$ (Section \ref{sec:xrays}) specifically, or indeed, high redshift proto-cluster environments more generally? Certainly, the apparent depolarization of the radio core (Figure \ref{fig:p_rm_ang} panel (\emph{d}); Section \ref{sec:results_pol}) is likely due to large RM dispersion in this region (e.g. \citealp{Burn1966,GW1966}). But only $\sim1$" (8.5 kpc) to the west of the radio core, we observe the fractional polarization increase to $\sim30\%$, with comparatively low net \absrm values of $\sim+100$ \radmm. This suggests that the radial extent of the gaseous structure responsible for the strong depolarization is $\lesssim8$ kpc, which is similar to that measured for gas bound to embedded cD-type galaxies, rather than the ICM proper (e.g. \citealp{Paolillo2002}). 

Since the X-ray data do not yield a detailed profile of $n_e$, we must resort to heuristics to understand the possible contribution of the ICM to the observed Faraday rotation measures, and by extension, reasonable values for the ICM magnetic field strength. To do so, we assume a simple model where the line-of-sight depth $D$ to each part of the radio source is identical (reasonable, given the radial extent of the ICM is a factor of three larger than that of the radio source). We approximate the electron density as being constant throughout this volume. We further assume that any given line of sight passes through a series of cells of size $l$ in which the orientation of the random magnetic field component (having strength $B_r$) is effectively independent of surrounding cells, and so likewise with the RM contribution of any sequence of cells falling along a given sight line. In such a case, the cluster gas will produce a dispersion in the RMs observed across independent sight-lines of \citep{Gaensler2001}:

\begin{equation}
\frac{\sigma_\text{RM}}{\text{rad m}^{-2}}=\frac{812}{\sqrt{3}}\Bigg(\frac{n_e}{\text{cm}^{-3}}\Bigg)\Bigg(\frac{B_r}{\muup G}\Bigg)\sqrt{\Bigg(\frac{D}{\text{kpc}}\Bigg)\Bigg(\frac{l}{\text{kpc}}\Bigg)}
\label{eqn:rmdispersion}
\end{equation}

\begin{figure} % Gen with spiderweb_X-band_hueint_alphablend_Lya.py
\centering
\includegraphics[width=0.45\textwidth]{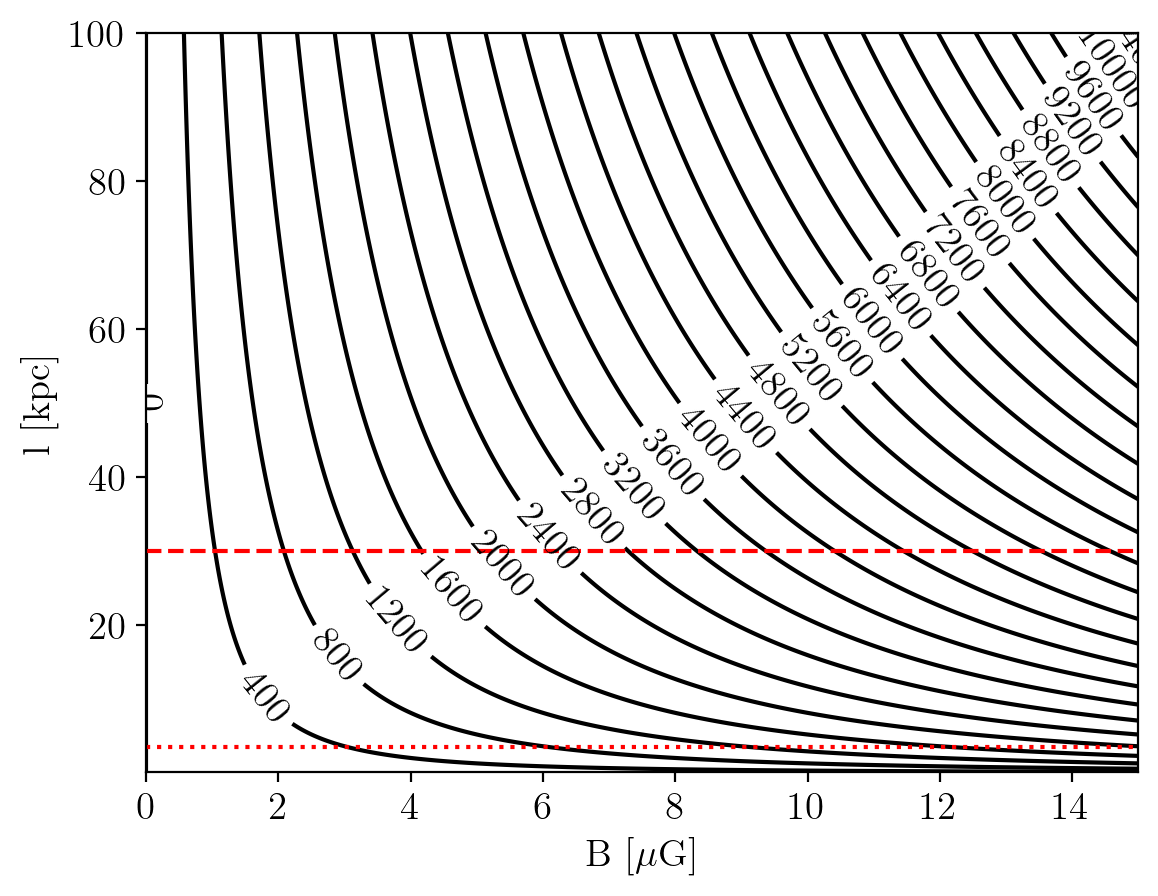} %plot_spiderweb_ICM_RM_contribution.py
\caption{Contour plot showing values of $\sigma_\text{RM}$ (in units of \radmm, in the source frame) as a function of the magnetic field strength $B$ and turbulent length scale $l$, derived from Equation \ref{eqn:rmdispersion} using fixed values of $n_e=1.5\times10^{-2}$ cm$^{-3}$ and $D=100$ kpc. The numbers on the contours are $\sigma_\text{RM}$ values in \radmm. The red dotted line is the FWHM of the common-resolution synthesized beam (0.5") at the distance of the cluster. The red dashed line is the approximate radial extent of the radio jets in the cluster gas.}
\label{fig:sigma_rm_vs_Bl}
\end{figure}

Taking $n_e=1.5\times10^{-2}$ cm$^{-3}$ and $D=100$ kpc, $\sigma_\text{RM}$ becomes a function of only $B$ and $l$, which we present as a contour plot in Figure \ref{fig:sigma_rm_vs_Bl}. We then use bootstrap re-sampling to estimate the actual RM dispersion in the radio jet, and its 90\% confidence interval, obtaining $\sigma_\text{RM,obs}=3120_{-920}^{+890}$ \radmm. From Figure \ref{fig:sigma_rm_vs_Bl}, supposing for the sake of argument that our estimate of $B_\text{hot}\sim9~\muup G$ in the jet-Ly$\alpha$ gas interaction region (Section \ref{sec:Lya_ancillary}) extends to the entire ICM, and furthermore that $D=100$ kpc is not a gross underestimate of the ICM scale, then $\sigma_\text{RM}$ values of the required magnitude can only be produced if the magnetic reversal scale satisfies $l\gtrsim25$ kpc, or 25\% of the cluster radius. This value would seem too large for an actively forming cluster. If the magnetic field strength is weaker than the assumed value, the magnetic reversal scale constraint increases, though we caution that a substantially stronger magnetic field could reduce the constraint derived above to more plausible values. Nevertheless, we conclude that RM structure observed on scales smaller than $\sim25$ kpc in the jets is probably not associated with mere RM dispersion from the ambient ICM, but rather gas in the vicinity of the radio jet itself.\\

So again we are left with the question: What are the most plausible range of values for $B$ and $l$ in the cluster gas? Consider the following. While we measure $|\text{RM}|=1100$ \radmm for the RM enhancement associated with the Ly$\alpha$ knot, the \absrm values upstream and downstream of the Ly$\alpha$ knot are either consistent with zero, or a factor of 4--12 times lower in absolute value. The same is true of the \absrm values found along most of the spine of the western jet before it bifurcates. In view of this, and in view of the other evidence we have supplied thus far which suggest that the RMs are predominantly generated in the vicinity of the jets (transverse RM gradients, higher \absrm values near the edges and ends of the jet than near the cluster centre and radio core), we propose a new `prior'\footnote{That is, `prior' in the Bayesian statistics sense of the term --- in this case, our \emph{a priori} beliefs about where RMs of a given magnitude are likely generated in the system} be adopted: Namely, that it is the relatively low \absrm values of $\mathcal{O}(200)$ \radmm that are generated by the ambient cluster gas, and over a path length some $100\text{kpc}/10\text{kpc}\approx10$ times greater than the Ly$\alpha$ RM enhancement. Since the gas pressure derived from our adopted volume-averaged electron density for the ICM (from \citet{Tozzi2022}; see Section \ref{sec:xrays}) is roughly equal to the pressure required to confine the Ly$\alpha$ knot (Section \ref{sec:Lya_ancillary}), we contend that the lower cluster-based RMs cannot be easily attributed to lower $n_e$ elsewhere in the cluster volume. The remaining possibilities are that (a) the magnetic field strength is at least a factor of 10 lower in the broader cluster gas than in the hot gas confining the Ly$\alpha$ knot, or that (b), numerous magnetic field reversals occur along any given sight line through the ICM, which will tend to reduce the observed RMs by a factor of $\sqrt{D/l}$. This latter effect is weak however, and would require a reversal scale that is smaller than the typical inner scale of turbulence in Galaxy clusters --- thought to be $\mathcal{O}(1)$ kpc; e.g. \citealt{Zhuravleva2018} --- to limit the RM produced by a field of only a few $\muup$G to only $\mathcal{O}(200)$ \radmm. For example, a pervasive $10 \muup$G field requires 1000 reversals along the line-of-sight (i.e. $l=0.1$ kpc) to produce sufficiently small \absrm values, and a $3 \muup$G field requires 100 reversals ($l=1$ kpc). While the inner scale of turbulence in high-$z$ proto-clusters might be expected to differ from that of nearby and mature galaxy clusters, small reversal scales on the order cited above would also give rise to a much patchier RM map than is observed. Thus, we claim that a realistic upper limit on the mean magnetic field strength in the ambient ICM of the Spiderweb system is $\mathcal{O}(1)~\muup$G.

\subsection{Magnetic fields in the interaction region}\label{sec:Bfieldsinteractionregion}

In Section \ref{sec:RMenhancement}, we discussed the presence of a hot gas component with $n_{e,hot}\sim0.015$ cm$^{-3}$, interacting with the radio jet and generating a local RM enhancement in the vicinity of a bright knot of Ly$\alpha$ emission. However, we have also now argued that the ICM is not primarily responsible for the observed RMs \emph{in general}, but rather that the RMs are generated in the immediate vicinity of the jet itself (Sections \ref{sec:extremeRMs}, \ref{sec:RMgradients}, and \ref{sec:icm}). We conclude that magnetic field strength calculations used in the specific case of the Ly$\alpha$ enhancement (Section \ref{sec:Lya_ancillary}) can be applied throughout the jet more broadly.

The \absrm values observed throughout the system are typically $\mathcal{O}(1000)$ \radmm, with values rising to $+11,100$ \radmm and $-4,400$ \radmm in the eastern and western jet respectively (Sections \ref{sec:extremeRMs} and \ref{sec:RMgradients}). Assuming that these RMs are generated over the same path length cited in Section \ref{sec:RMenhancement} (10 kpc), and that the Faraday-rotating gas has $n_{e,hot}\sim0.015$ cm$^{-3}$, we estimate that the magnetic field strength in the immediate vicinity of the jet is typically $B_\text{jet}\sim10~\muup G$, rising to $\sim35~\muup G$ around the southern terminus of the western jet, and $\sim90~\muup G$ around the hotspot complex of the eastern jet (all values rounded to the nearest 5 $\muup G$). We note the similarity of these values to those derived by \citet{Carilli2022} --- 50--$70\muup$G, corresponding to an emission-weighted average over the lobe volume --- from their analysis of inverse-Compton scattering in the system. 

\section{Discussion}\label{sec:discussion}

We have presented the most detailed map of the magnetic field structure in and around a high-$z$ radio galaxy to date. The RM, fractional polarization, and projected magnetic field orientation all show smooth changes in structure on scales larger than the synthesized beamwidth. The projected magnetic field orientation is oriented almost exactly along the axis of the jet along its entire length, indicating a substantial poloidal component to the fields. Our detection of large-scale cross-jet RM gradients highlight a coherent toroidal field component (discussed in more detail below). The ratio of the ordered to random components in the plane-of-sky-projected magnetic field is of order unity. The magnetic fields in the radio jet are therefore well-ordered on scales as small as $\sim$a few kpc, up to the size scale of the radio jet itself ($\sim$a few tens of kpc), only $\sim3$ Gyr after the Big Bang.\\

Contrary to previous work (e.g. \citealp{Pentericci1997}) in which large \absrm  values observed towards the radio jet were assumed to arise in the ambient cluster gas, we have presented new evidence to suggest the Faraday rotation principally occurs in the immediate vicinity of the jet, and in the gas interacting with it, as follows. The most extreme \absrm values in the system do not occur towards the central regions of the proto-cluster, but towards the eastern and western terminal points of the radio jet, and the edges of the jet (Table \ref{table2}; also Figure \ref{fig:p_rm_ang} (b)). The eastern hotspot complex shows RM interfaces (that is, large changes in RM over angular scales smaller than the synthesized beamwidth; Figure \ref{fig:p_rm_ang} (b)), which are correlated with structure in the total intensity emission (Figure \ref{fig:s-x-ka-total-int}). There are transverse gradients in RM across the western jet (Figure \ref{fig:rm_gradients_west_jet}), which cross zero \radmm and thereby indicate a change in the line-of-sight-projected magnetic field direction. This orientation-flip is spatially coincident with the spine of the radio jet. We also observe a spatial correlation between a bright Ly$\alpha$ knot, the deviation and bifurcation of the radio jet in total intensity, and a local enhancement in the \absrm of $\sim1100$ \radmm (Figure \ref{fig:fd_lya_olay}). Finally, we argued that the magnetic field strength in the cluster ICM is low, and cannot explain RM structure observed on $\sim$arcsecond scales within the jet (Section \ref{sec:icm}). Taken together with some key findings of \citet{Carilli2022} --- i.e. that the synchrotron plasma is inverse-Compton-up-scattering the cosmic microwave background, and that the radio source has a hybrid FRI/II morphology which appears to be coupled to its one-sided displacement of Ly$\alpha$-emitting gas in the system --- a consistent picture emerges of a radio jet undergoing vigorous interaction with the surrounding gaseous environment, which is likely still in the process of falling into the system and equilibrating. The strong jet-gas interaction signatures may mean that the radio jet drives outflows observed in the system, rather than radiation pressure and winds from the central AGN, though these mechanisms are all energetically capable (Seymour 2012, Nesvadba 2017). It is significant that this is observed at $z\sim2$, where the importance of `quasar-mode' AGN feedback wanes, the importance of `jet-mode' or `radio-mode' feedback begins to dominate, and galaxies transition from more active to passive modes of evolution (Section \ref{sec:intro}).\\

The RM gradients identified in Section \ref{sec:RMgradients} increase anti-clockwise when viewed down the jet from the central AGN, presumably tracing toroidal magnetic field components with a clockwise orientation. Combined with our results that the sky-projected magnetic field lies broadly parallel to the jet along its length (Section \ref{sec:detailedmapB}), and that this field is well-ordered (Section \ref{sec:detailedmapB}), it is likely that the overall magnetic field structure of the jet is helical. \citet{Christodoulou2016} and \citet{Gabuzda2018,Gabuzda2018b} have cited a statistical preference for clockwise toroidal magnetic field components observed at decaparsec--kiloparsec scales (coupled with counterclockwise toroidal magnetic field components observed at parsec scales; \citealp{Contopoulos2009}) as evidence of a predicted manifestation of the Poynting-Robertson `cosmic battery' \citep{Christodoulou2016} generating nested helical magnetic fields (e.g. see Figure \ref{fig:toy_model}; also Figure 1 of \citealp{Contopoulos2009} and Figure 2 of \citealp{Gabuzda2018}). The sense of the RM gradient we detect is consistent with this picture. The results imply that the net current is flowing outward in the cylindrical region enclosed by the magnetic fields that dominate the RM signal. The cosmic battery is a means of generating strong, ordered, large-scale magnetic fields in the accretion disks of AGN, through which radio jets can be seeded with helical magnetic fields, and which in turn may transport magnetic flux from AGN into the broader universe. We propose that the Spiderweb radio galaxy may show the cosmic battery in action.\\

The hierarchy of estimated magnetic field strengths in the system are of further relevance to the findings just discussed. The magnetic field strength is $\sim90~\muup$G the terminal regions of the jets, where interaction with the surrounding gas is obviously maximal. The field strength drops to 50--$70\muup$G in the radio lobe volume \citep{Carilli2022}, then drops further to typically $10\muup$G in the gas interacting with the jets, and finally drops again to perhaps $1\muup$G in the ambient ICM. We note that the magnetic pressure in and around the jet does not appear to dominate the jet's dynamical interactions, but it may not be entirely negligible given the substantial uncertainties. Considering (a) the cascade of decreasing magnetic field strengths as we go from the lobes, to gas interacting with the lobes, to the ambient ICM; (b) the proposed operation of a cosmic battery in the jets; (c) that the radio galaxy is embedded in a proto-cluster environment at high redshift; and (d) the evidence for strong jet-gas interactions throughout the system, we further propose that the Spiderweb radio jet may be in the process of magnetising the cluster gas, at a redshift which is relevant to broader questions around how the Universe became pervasively magnetized, as it appears to be (e.g. see \citealp{NV2010,Chen2015,Govoni2019,Vazza2021}). Recent results from the IllustrisTNG simulations \citep{AG2021} appear to back this up, not only demonstrating the probable cosmological importance of AGN-driven magnetization of the Universe generally, but also the escalating importance of these processes at $z\sim2$ specifically --- again placing the $z=2.16$ Spiderweb system near the dawn of an important cosmic epoch as far as these questions go.\\ 

Finally, our clear detections of (i) \absrm values up to 10,000 \radmm, (ii) kpc-scale transverse RM gradients, and (iii) radio-mode feedback, particularly in the form an \absrm enhancement co-located with a bright Ly$\alpha$ knot and jet path deviation, are all novel for high-$z$ radio galaxies, and demonstrate the value of modern spectropolarimetric analysis in this regime. In making these observations however, even the Jansky VLA is near to its effective limits of sensitivity, resolution, and object-frame $\lambda^2$ coverage, such that it is probably impossible to study statistical samples of high-$z$ radio galaxies in comparable detail at the present time. Such studies must wait for the combination of sensitivity, resolving power, and high-frequency bandwidth coverage that will be provided by the Next Generation VLA \citep{Murphy2018}, and the Square Kilometer Array (\citealp{Heald2020}, and references therein). In the interim, we plan to test the broader importance of our findings by targeting a select few other high-$z$ radio galaxies using spectropolarimetric Jansky VLA observations, including 4C41.17 at $z=3.6$ (e.g. see \citealp{MD2008}).

\begin{acknowledgements}
We thank the anonymous referee for their time and helpful comments, which have significantly improved the content and presentation of this paper. This work employed VLA data from projects 19A-024 and 20B-488. The National Radio Astronomy Observatory is a facility of the National Science Foundation operated under cooperative agreement by Associated Universities, Inc. ~C.~S.~A. is a Jansky Fellow of the NRAO. ~C.~L.~C. acknowledges support through CXC grant G09-2-1103X. ~S.~B. acknowledges financial support from the INFN INDARK grant.
~P.~T. and ~S.~B. acknowledge financial support from the agreement ASI-INAF n.2017-14-H.0. ~A.~S. is supported by the ERC-StG ‘ClustersXCosmo’ grant agreement 716762, by
the FARE-MIUR grant ‘ClustersXEuclid’ R165SBKTMA, and by INFN InDark Grant.
\end{acknowledgements}

\bibliographystyle{pasa-mnras}
\bibliography{bibliography}

\newcommand{\noop}[1]{}
\begin{thebibliography}{}
\makeatletter
\relax
\def\mn@urlcharsother{\let\do\@makeother \do\$\do\&\do\#\do\^\do\_\do\%\do\~}
\definecolor{darkblue}{rgb}{0,0,0.597656}
\def\mndoi{\begingroup\mn@urlcharsother \@ifnextchar [ {\mndoi@} {\mndoi@[]}}
\def\mndoi@[#1]#2{\def\@tempa{#1}\ifx\@tempa\@empty \href
  {http://dx.doi.org/#2} {\textcolor{darkblue}{doi:#2}}\else \href
  {http://dx.doi.org/#2} {\textcolor{darkblue}{#1}}\fi \endgroup}
\def\mn@eprint#1#2{\mn@eprint@#1:#2::\@nil}
\def\mn@eprint@arXiv#1{\href {http://arxiv.org/abs/#1} {{\tt arXiv:#1}}}
\def\mn@eprint@dblp#1{\href {http://dblp.uni-trier.de/rec/bibtex/#1.xml}
  {dblp:#1}}
\def\mn@eprint@#1:#2:#3:#4\@nil{\def\@tempa {#1}\def\@tempb {#2}\def\@tempc
  {#3}\ifx \@tempc \@empty \let \@tempc \@tempb \let \@tempb \@tempa \fi \ifx
  \@tempb \@empty \def\@tempb {arXiv}\fi \@ifundefined
  {mn@eprint@\@tempb}{\@tempb:\@tempc}{\expandafter \expandafter \csname
  mn@eprint@\@tempb\endcsname \expandafter{\@tempc}}}

\bibitem[\protect\citeauthoryear{{Abdo} et~al.,}{{Abdo}
  et~al.}{2010}]{Abdo2010}
{Abdo} A.~A.,  et~al., 2010, \mndoi [Science] {10.1126/science.1184656}, \href
  {http://adsabs.harvard.edu/abs/2010Sci...328..725A} {328, 725}

\bibitem[\protect\citeauthoryear{Anderson}{Anderson}{2016}]{Anderson2016b}
Anderson C.,  2016, PhD thesis, The University of Sydney

\bibitem[\protect\citeauthoryear{{Anderson}, {Gaensler}, {Feain}  \&
  {Franzen}}{{Anderson} et~al.}{2015}]{Anderson2015}
{Anderson} C.~S.,  {Gaensler} B.~M.,  {Feain} I.~J.,   {Franzen} T.~M.~O.,
  2015, \mndoi [\apj] {10.1088/0004-637X/815/1/49}, \href
  {http://adsabs.harvard.edu/abs/2015ApJ...815...49A} {815, 49}

\bibitem[\protect\citeauthoryear{{Anderson}, {Gaensler}  \& {Feain}}{{Anderson}
  et~al.}{2016}]{Anderson2016}
{Anderson} C.~S.,  {Gaensler} B.~M.,   {Feain} I.~J.,  2016, \mndoi [\apj]
  {10.3847/0004-637X/825/1/59}, \href
  {http://adsabs.harvard.edu/abs/2016ApJ...825...59A} {825, 59}

\bibitem[\protect\citeauthoryear{Anderson et~al.,}{Anderson
  et~al.}{2018a}]{Anderson2018b}
Anderson C.~S.,  et~al., 2018a, \mndoi [Galaxies] {10.3390/galaxies6040127}, 6

\bibitem[\protect\citeauthoryear{{Anderson}, {Gaensler}, {Heald}, {O'Sullivan},
  {Kaczmarek}  \& {Feain}}{{Anderson} et~al.}{2018b}]{Anderson2018}
{Anderson} C.~S.,  {Gaensler} B.~M.,  {Heald} G.~H.,  {O'Sullivan} S.~P.,
  {Kaczmarek} J.~F.,   {Feain} I.~J.,  2018b, \mndoi [\apj]
  {10.3847/1538-4357/aaaec0}, \href
  {http://adsabs.harvard.edu/abs/2018ApJ...855...41A} {855, 41}

\bibitem[\protect\citeauthoryear{{Anderson} et~al.,}{{Anderson}
  et~al.}{2021}]{Anderson2021}
{Anderson} C.~S.,  et~al., 2021, \mndoi [\pasa] {10.1017/pasa.2021.4}, \href
  {https://ui.adsabs.harvard.edu/abs/2021PASA...38...20A} {38, e020}

\bibitem[\protect\citeauthoryear{{Ar{\'a}mburo-Garc{\'\i}a}, {Bondarenko},
  {Boyarsky}, {Nelson}, {Pillepich}  \& {Sokolenko}}{{Ar{\'a}mburo-Garc{\'\i}a}
  et~al.}{2021}]{AG2021}
{Ar{\'a}mburo-Garc{\'\i}a} A.,  {Bondarenko} K.,  {Boyarsky} A.,  {Nelson} D.,
  {Pillepich} A.,   {Sokolenko} A.,  2021, \mndoi [\mnras]
  {10.1093/mnras/stab1632}, \href
  {https://ui.adsabs.harvard.edu/abs/2021MNRAS.505.5038A} {505, 5038}

\bibitem[\protect\citeauthoryear{{Attridge}, {Roberts}  \& {Wardle}}{{Attridge}
  et~al.}{1999}]{Attridge1999}
{Attridge} J.~M.,  {Roberts} D.~H.,   {Wardle} J.~F.~C.,  1999, \mndoi [\apjl]
  {10.1086/312078}, \href {http://adsabs.harvard.edu/abs/1999ApJ...518L..87A}
  {518, L87}

\bibitem[\protect\citeauthoryear{{Bertin}, {Mellier}, {Radovich}, {Missonnier},
  {Didelon}  \& {Morin}}{{Bertin} et~al.}{2002}]{Bertin2002}
{Bertin} E.,  {Mellier} Y.,  {Radovich} M.,  {Missonnier} G.,  {Didelon} P.,
  {Morin} B.,  2002, in {Bohlender} D.~A.,  {Durand} D.,   {Handley} T.~H.,
  eds,  Astronomical Society of the Pacific Conference Series Vol. 281,
  Astronomical Data Analysis Software and Systems XI. p.~228

\bibitem[\protect\citeauthoryear{{Bonafede}, {Feretti}, {Murgia}, {Govoni},
  {Giovannini}, {Dallacasa}, {Dolag}  \& {Taylor}}{{Bonafede}
  et~al.}{2010}]{Bonafede2010}
{Bonafede} A.,  {Feretti} L.,  {Murgia} M.,  {Govoni} F.,  {Giovannini} G.,
  {Dallacasa} D.,  {Dolag} K.,   {Taylor} G.~B.,  2010, \mndoi [\aap]
  {10.1051/0004-6361/200913696}, \href
  {http://adsabs.harvard.edu/abs/2010A%26A...513A..30B} {513, A30}

\bibitem[\protect\citeauthoryear{{Brammer} et~al.,}{{Brammer}
  et~al.}{2009}]{Brammer2009}
{Brammer} G.~B.,  et~al., 2009, \mndoi [\apjl] {10.1088/0004-637X/706/1/L173},
  \href {https://ui.adsabs.harvard.edu/abs/2009ApJ...706L.173B} {706, L173}

\bibitem[\protect\citeauthoryear{{Brentjens} \& {de Bruyn}}{{Brentjens} \& {de
  Bruyn}}{2005}]{BdB2005}
{Brentjens} M.~A.,  {de Bruyn} A.~G.,  2005, \mndoi [\aap]
  {10.1051/0004-6361:20052990}, \href
  {http://adsabs.harvard.edu/abs/2005A%26A...441.1217B} {441, 1217}

\bibitem[\protect\citeauthoryear{{Briggs}}{{Briggs}}{1995}]{Briggs1995}
{Briggs} D.~S.,  1995, in American Astronomical Society Meeting Abstracts.
  p.~1444

\bibitem[\protect\citeauthoryear{{Burn}}{{Burn}}{1966}]{Burn1966}
{Burn} B.~J.,  1966, \mndoi [\mnras] {10.1093/mnras/133.1.67}, \href
  {http://adsabs.harvard.edu/abs/1966MNRAS.133...67B} {133, 67}

\bibitem[\protect\citeauthoryear{{Carilli} \& {Taylor}}{{Carilli} \&
  {Taylor}}{2002}]{CT2002}
{Carilli} C.~L.,  {Taylor} G.~B.,  2002, \mndoi [\araa]
  {10.1146/annurev.astro.40.060401.093852}, \href
  {http://adsabs.harvard.edu/abs/2002ARA%26A..40..319C} {40, 319}

\bibitem[\protect\citeauthoryear{{Carilli}, {R{\"o}ttgering}, {van Ojik},
  {Miley}  \& {van Breugel}}{{Carilli} et~al.}{1997}]{Carilli1997}
{Carilli} C.~L.,  {R{\"o}ttgering} H.~J.~A.,  {van Ojik} R.,  {Miley} G.~K.,
  {van Breugel} W.~J.~M.,  1997, \mndoi [\apjs] {10.1086/312973}, \href
  {https://ui.adsabs.harvard.edu/abs/1997ApJS..109....1C} {109, 1}

\bibitem[\protect\citeauthoryear{{Carilli}, {Harris}, {Pentericci},
  {R{\"o}ttgering}, {Miley}  \& {Bremer}}{{Carilli} et~al.}{1998}]{Carilli1998}
{Carilli} C.~L.,  {Harris} D.~E.,  {Pentericci} L.,  {R{\"o}ttgering} H.~J.~A.,
   {Miley} G.~K.,   {Bremer} M.~N.,  1998, \mndoi [\apjl] {10.1086/311199},
  \href {https://ui.adsabs.harvard.edu/abs/1998ApJ...494L.143C} {494, L143}

\bibitem[\protect\citeauthoryear{{Carilli}, {Harris}, {Pentericci},
  {R{\"o}ttgering}, {Miley}, {Kurk}  \& {van Breugel}}{{Carilli}
  et~al.}{2002}]{Carilli2002}
{Carilli} C.~L.,  {Harris} D.~E.,  {Pentericci} L.,  {R{\"o}ttgering} H.~J.~A.,
   {Miley} G.~K.,  {Kurk} J.~D.,   {van Breugel} W.,  2002, \mndoi [\apj]
  {10.1086/338669}, \href
  {https://ui.adsabs.harvard.edu/abs/2002ApJ...567..781C} {567, 781}

\bibitem[\protect\citeauthoryear{{Carilli} et~al.,}{{Carilli}
  et~al.}{2022}]{Carilli2022}
{Carilli} C.~L.,  et~al., 2022, \mndoi [\apj] {10.3847/1538-4357/ac55a0}, \href
  {https://ui.adsabs.harvard.edu/abs/2022ApJ...928...59C} {928, 59}

\bibitem[\protect\citeauthoryear{{Chambers}, {Miley}  \& {van
  Breugel}}{{Chambers} et~al.}{1990}]{Chambers1990}
{Chambers} K.~C.,  {Miley} G.~K.,   {van Breugel} W.~J.~M.,  1990, \mndoi
  [\apj] {10.1086/169316}, \href
  {https://ui.adsabs.harvard.edu/abs/1990ApJ...363...21C} {363, 21}

\bibitem[\protect\citeauthoryear{Chen, Buckley  \& Ferrer}{Chen
  et~al.}{2015}]{Chen2015}
Chen W.,  Buckley J.~H.,   Ferrer F.,  2015, \mndoi [Phys. Rev. Lett.]
  {10.1103/PhysRevLett.115.211103}, 115, 211103

\bibitem[\protect\citeauthoryear{{Christodoulou}, {Gabuzda}, {Knuettel},
  {Contopoulos}, {Kazanas}  \& {Coughlan}}{{Christodoulou}
  et~al.}{2016}]{Christodoulou2016}
{Christodoulou} D.~M.,  {Gabuzda} D.~C.,  {Knuettel} S.,  {Contopoulos} I.,
  {Kazanas} D.,   {Coughlan} C.~P.,  2016, \mndoi [\aap]
  {10.1051/0004-6361/201527448}, \href
  {https://ui.adsabs.harvard.edu/abs/2016A&A...591A..61C} {591, A61}

\bibitem[\protect\citeauthoryear{{Contopoulos}, {Christodoulou}, {Kazanas}  \&
  {Gabuzda}}{{Contopoulos} et~al.}{2009}]{Contopoulos2009}
{Contopoulos} I.,  {Christodoulou} D.~M.,  {Kazanas} D.,   {Gabuzda} D.~C.,
  2009, \mndoi [\apjl] {10.1088/0004-637X/702/2/L148}, \href
  {https://ui.adsabs.harvard.edu/abs/2009ApJ...702L.148C} {702, L148}

\bibitem[\protect\citeauthoryear{{Conway}, {Haves}, {Kronberg}, {Stannard},
  {Vallee}  \& {Wardle}}{{Conway} et~al.}{1974}]{Conway1974}
{Conway} R.~G.,  {Haves} P.,  {Kronberg} P.~P.,  {Stannard} D.,  {Vallee}
  J.~P.,   {Wardle} J.~F.~C.,  1974, \mndoi [\mnras] {10.1093/mnras/168.1.137},
  \href {http://adsabs.harvard.edu/abs/1974MNRAS.168..137C} {168}

\bibitem[\protect\citeauthoryear{{Cooper} \& {Price}}{{Cooper} \&
  {Price}}{1962}]{CP1962}
{Cooper} B.~F.~C.,  {Price} R.~M.,  1962, \mndoi [\nat] {10.1038/196761a0},
  \href {https://ui.adsabs.harvard.edu/abs/1962Natur.196..761C} {196, 761}

\bibitem[\protect\citeauthoryear{{Croft} et~al.,}{{Croft}
  et~al.}{2006}]{Croft2006}
{Croft} S.,  et~al., 2006, \mndoi [\apj] {10.1086/505526}, \href
  {https://ui.adsabs.harvard.edu/abs/2006ApJ...647.1040C} {647, 1040}

\bibitem[\protect\citeauthoryear{{Croton} et~al.,}{{Croton}
  et~al.}{2006}]{Croton2006}
{Croton} D.~J.,  et~al., 2006, \mndoi [\mnras]
  {10.1111/j.1365-2966.2005.09675.x}, \href
  {http://adsabs.harvard.edu/abs/2006MNRAS.365...11C} {365, 11}

\bibitem[\protect\citeauthoryear{{Dabbech}, {Onose}, {Abdulaziz}, {Perley},
  {Smirnov}  \& {Wiaux}}{{Dabbech} et~al.}{2018}]{Dabbech2018}
{Dabbech} A.,  {Onose} A.,  {Abdulaziz} A.,  {Perley} R.~A.,  {Smirnov} O.~M.,
   {Wiaux} Y.,  2018, \mndoi [\mnras] {10.1093/mnras/sty372}, \href
  {https://ui.adsabs.harvard.edu/abs/2018MNRAS.476.2853D} {476, 2853}

\bibitem[\protect\citeauthoryear{{Donnert}, {Vazza}, {Br{\"u}ggen}  \&
  {ZuHone}}{{Donnert} et~al.}{2018}]{Donnert2018}
{Donnert} J.,  {Vazza} F.,  {Br{\"u}ggen} M.,   {ZuHone} J.,  2018, \mndoi
  [\ssr] {10.1007/s11214-018-0556-8}, \href
  {https://ui.adsabs.harvard.edu/abs/2018SSRv..214..122D} {214, 122}

\bibitem[\protect\citeauthoryear{Dreher, Carilli  \& Perley}{Dreher
  et~al.}{1987}]{Dreher87}
Dreher J.,  Carilli C.,   Perley R.,  1987, \mndoi [The Astrophysical Journal]
  {10.1086/165229}, 316, 611

\bibitem[\protect\citeauthoryear{{Eichmann}, {Rachen}, {Merten}, {van Vliet}
  \& {Becker Tjus}}{{Eichmann} et~al.}{2018}]{Eichmann_2018}
{Eichmann} B.,  {Rachen} J.~P.,  {Merten} L.,  {van Vliet} A.,   {Becker Tjus}
  J.,  2018, \mndoi [jcap] {10.1088/1475-7516/2018/02/036}, \href
  {https://ui.adsabs.harvard.edu/abs/2018JCAP...02..036E} {2018, 036}

\bibitem[\protect\citeauthoryear{{Emonts} et~al.,}{{Emonts}
  et~al.}{2016}]{Emonts2016}
{Emonts} B.~H.~C.,  et~al., 2016, \mndoi [Science] {10.1126/science.aag0512},
  \href {https://ui.adsabs.harvard.edu/abs/2016Sci...354.1128E} {354, 1128}

\bibitem[\protect\citeauthoryear{{Event Horizon Telescope Collaboration}
  et~al.,}{{Event Horizon Telescope Collaboration} et~al.}{2021}]{EHTC2021}
{Event Horizon Telescope Collaboration} et~al., 2021, \mndoi [\apjl]
  {10.3847/2041-8213/abe4de}, \href
  {https://ui.adsabs.harvard.edu/abs/2021ApJ...910L..13E} {910, L13}

\bibitem[\protect\citeauthoryear{{Falkendal} et~al.,}{{Falkendal}
  et~al.}{2019}]{Falkendal2019}
{Falkendal} T.,  et~al., 2019, \mndoi [\aap] {10.1051/0004-6361/201732485},
  \href {https://ui.adsabs.harvard.edu/abs/2019A&A...621A..27F} {621, A27}

\bibitem[\protect\citeauthoryear{{Falkendal}, {Lehnert}, {Vernet}, {De Breuck}
  \& {Wang}}{{Falkendal} et~al.}{2021}]{Falkendal2021}
{Falkendal} T.,  {Lehnert} M.~D.,  {Vernet} J.,  {De Breuck} C.,   {Wang} W.,
  2021, \mndoi [\aap] {10.1051/0004-6361/201935237}, \href
  {https://ui.adsabs.harvard.edu/abs/2021A&A...645A.120F} {645, A120}

\bibitem[\protect\citeauthoryear{{Farnsworth}, {Rudnick}  \&
  {Brown}}{{Farnsworth} et~al.}{2011}]{Farnsworth2011}
{Farnsworth} D.,  {Rudnick} L.,   {Brown} S.,  2011, \mndoi [\aj]
  {10.1088/0004-6256/141/6/191}, \href
  {http://adsabs.harvard.edu/abs/2011AJ....141..191F} {141, 191}

\bibitem[\protect\citeauthoryear{{F{\"o}rster Schreiber} \&
  {Wuyts}}{{F{\"o}rster Schreiber} \& {Wuyts}}{2020}]{Schreiber2020}
{F{\"o}rster Schreiber} N.~M.,  {Wuyts} S.,  2020, \mndoi [\araa]
  {10.1146/annurev-astro-032620-021910}, \href
  {https://ui.adsabs.harvard.edu/abs/2020ARA&A..58..661F} {58, 661}

\bibitem[\protect\citeauthoryear{{Fragile}, {Anninos}, {Croft}, {Lacy}  \&
  {Witry}}{{Fragile} et~al.}{2017}]{Fragile2017}
{Fragile} P.~C.,  {Anninos} P.,  {Croft} S.,  {Lacy} M.,   {Witry} J. W.~L.,
  2017, \mndoi [\apj] {10.3847/1538-4357/aa95c6}, \href
  {https://ui.adsabs.harvard.edu/abs/2017ApJ...850..171F} {850, 171}

\bibitem[\protect\citeauthoryear{{Furlanetto} \& {Loeb}}{{Furlanetto} \&
  {Loeb}}{2001}]{FL2001}
{Furlanetto} S.~R.,  {Loeb} A.,  2001, \mndoi [\apj] {10.1086/321630}, \href
  {http://adsabs.harvard.edu/abs/2001ApJ...556..619F} {556, 619}

\bibitem[\protect\citeauthoryear{{Gabuzda}}{{Gabuzda}}{2018}]{Gabuzda2018}
{Gabuzda} D.,  2018, \mndoi [Galaxies] {10.3390/galaxies7010005}, \href
  {https://ui.adsabs.harvard.edu/abs/2018Galax...7....5G} {7, 5}

\bibitem[\protect\citeauthoryear{{Gabuzda}, {Nagle}  \& {Roche}}{{Gabuzda}
  et~al.}{2018}]{Gabuzda2018b}
{Gabuzda} D.~C.,  {Nagle} M.,   {Roche} N.,  2018, \mndoi [\aap]
  {10.1051/0004-6361/201732136}, \href
  {https://ui.adsabs.harvard.edu/abs/2018A&A...612A..67G} {612, A67}

\bibitem[\protect\citeauthoryear{{Gaensler}, {Dickey}, {McClure-Griffiths},
  {Green}, {Wieringa}  \& {Haynes}}{{Gaensler} et~al.}{2001}]{Gaensler2001}
{Gaensler} B.~M.,  {Dickey} J.~M.,  {McClure-Griffiths} N.~M.,  {Green} A.~J.,
  {Wieringa} M.~H.,   {Haynes} R.~F.,  2001, \mndoi [\apj] {10.1086/319468},
  \href {https://ui.adsabs.harvard.edu/abs/2001ApJ...549..959G} {549, 959}

\bibitem[\protect\citeauthoryear{{Gaensler} et~al.,}{{Gaensler}
  et~al.}{2015}]{Gaensler2015}
{Gaensler} B.,  et~al., 2015, Advancing Astrophysics with the Square Kilometre
  Array (AASKA14), \href {http://adsabs.harvard.edu/abs/2015aska.confE.103G}
  {p.~103}

\bibitem[\protect\citeauthoryear{{Gaibler}, {Khochfar}, {Krause}  \&
  {Silk}}{{Gaibler} et~al.}{2012}]{Gaibler2012}
{Gaibler} V.,  {Khochfar} S.,  {Krause} M.,   {Silk} J.,  2012, \mndoi [\mnras]
  {10.1111/j.1365-2966.2012.21479.x}, \href
  {https://ui.adsabs.harvard.edu/abs/2012MNRAS.425..438G} {425, 438}

\bibitem[\protect\citeauthoryear{{Gardner} \& {Whiteoak}}{{Gardner} \&
  {Whiteoak}}{1966}]{GW1966}
{Gardner} F.~F.,  {Whiteoak} J.~B.,  1966, \mndoi [\araa]
  {10.1146/annurev.aa.04.090166.001333}, \href
  {http://adsabs.harvard.edu/abs/1966ARA%26A...4..245G} {4, 245}

\bibitem[\protect\citeauthoryear{{Gaspari} et~al.,}{{Gaspari}
  et~al.}{2019}]{Gaspari2019}
{Gaspari} M.,  et~al., 2019, \mndoi [\apj] {10.3847/1538-4357/ab3c5d}, \href
  {https://ui.adsabs.harvard.edu/abs/2019ApJ...884..169G} {884, 169}

\bibitem[\protect\citeauthoryear{{Govoni} et~al.,}{{Govoni}
  et~al.}{2019}]{Govoni2019}
{Govoni} F.,  et~al., 2019, \mndoi [Science] {10.1126/science.aat7500}, \href
  {https://ui.adsabs.harvard.edu/abs/2019Sci...364..981G} {364, 981}

\bibitem[\protect\citeauthoryear{{Gullberg} et~al.,}{{Gullberg}
  et~al.}{2016}]{Gullberg2016}
{Gullberg} B.,  et~al., 2016, \mndoi [\aap] {10.1051/0004-6361/201527647},
  \href {https://ui.adsabs.harvard.edu/abs/2016A&A...591A..73G} {591, A73}

\bibitem[\protect\citeauthoryear{{Hardcastle} \& {Croston}}{{Hardcastle} \&
  {Croston}}{2020}]{HC2020}
{Hardcastle} M.~J.,  {Croston} J.~H.,  2020, \mndoi [nar]
  {10.1016/j.newar.2020.101539}, \href
  {https://ui.adsabs.harvard.edu/abs/2020NewAR..8801539H} {88, 101539}

\bibitem[\protect\citeauthoryear{{Hardcastle}, {Cheung}, {Feain}  \&
  {Stawarz}}{{Hardcastle} et~al.}{2009}]{Hardcastle2009}
{Hardcastle} M.~J.,  {Cheung} C.~C.,  {Feain} I.~J.,   {Stawarz} {\L}.,  2009,
  \mndoi [\mnras] {10.1111/j.1365-2966.2008.14265.x}, \href
  {http://adsabs.harvard.edu/abs/2009MNRAS.393.1041H} {393, 1041}

\bibitem[\protect\citeauthoryear{Hatch et~al.,}{Hatch
  et~al.}{2014}]{Hatch_2014}
Hatch N.~A.,  et~al., 2014, \mndoi [Monthly Notices of the Royal Astronomical
  Society] {10.1093/mnras/stu1725}, 445, 280

\bibitem[\protect\citeauthoryear{{Heald}, {Braun}  \& {Edmonds}}{{Heald}
  et~al.}{2009}]{Heald2009}
{Heald} G.,  {Braun} R.,   {Edmonds} R.,  2009, \mndoi [\aap]
  {10.1051/0004-6361/200912240}, \href
  {http://adsabs.harvard.edu/abs/2009A%26A...503..409H} {503, 409}

\bibitem[\protect\citeauthoryear{{Heald} et~al.,}{{Heald}
  et~al.}{2020}]{Heald2020}
{Heald} G.,  et~al., 2020, \mndoi [Galaxies] {10.3390/galaxies8030053}, \href
  {https://ui.adsabs.harvard.edu/abs/2020Galax...8...53H} {8, 53}

\bibitem[\protect\citeauthoryear{Hodges-Kluck, Gallo, Ghisellini, Haardt, Wu
  \& Ciardi}{Hodges-Kluck et~al.}{2021}]{Hodges-Kluck2021}
Hodges-Kluck E.,  Gallo E.,  Ghisellini G.,  Haardt F.,  Wu J.,   Ciardi B.,
  2021, Proof of CMB-driven X-ray brightening of high-z radio galaxies
  (\mn@eprint {arXiv} {2105.03467})

\bibitem[\protect\citeauthoryear{{Hovatta}, {Lister}, {Aller}, {Aller},
  {Homan}, {Kovalev}, {Pushkarev}  \& {Savolainen}}{{Hovatta}
  et~al.}{2012}]{Hovatta2012}
{Hovatta} T.,  {Lister} M.~L.,  {Aller} M.~F.,  {Aller} H.~D.,  {Homan} D.~C.,
  {Kovalev} Y.~Y.,  {Pushkarev} A.~B.,   {Savolainen} T.,  2012, \mndoi [\aj]
  {10.1088/0004-6256/144/4/105}, \href
  {http://adsabs.harvard.edu/abs/2012AJ....144..105H} {144, 105}

\bibitem[\protect\citeauthoryear{{Hughes}, {Aller}  \& {Aller}}{{Hughes}
  et~al.}{1989a}]{HAA1989a}
{Hughes} P.~A.,  {Aller} H.~D.,   {Aller} M.~F.,  1989a, \mndoi [\apj]
  {10.1086/167471}, \href
  {https://ui.adsabs.harvard.edu/abs/1989ApJ...341...54H} {341, 54}

\bibitem[\protect\citeauthoryear{{Hughes}, {Aller}  \& {Aller}}{{Hughes}
  et~al.}{1989b}]{HAA1989b}
{Hughes} P.~A.,  {Aller} H.~D.,   {Aller} M.~F.,  1989b, \mndoi [\apj]
  {10.1086/167472}, \href
  {https://ui.adsabs.harvard.edu/abs/1989ApJ...341...68H} {341, 68}

\bibitem[\protect\citeauthoryear{{Hughes}, {Aller}  \& {Aller}}{{Hughes}
  et~al.}{1991}]{HAA1991}
{Hughes} P.~A.,  {Aller} H.~D.,   {Aller} M.~F.,  1991, \mndoi [\apj]
  {10.1086/170096}, \href
  {https://ui.adsabs.harvard.edu/abs/1991ApJ...374...57H} {374, 57}

\bibitem[\protect\citeauthoryear{{Hutschenreuter} \&
  {En{\ss}lin}}{{Hutschenreuter} \& {En{\ss}lin}}{2020}]{HE2020}
{Hutschenreuter} S.,  {En{\ss}lin} T.~A.,  2020, \mndoi [\aap]
  {10.1051/0004-6361/201935479}, \href
  {https://ui.adsabs.harvard.edu/abs/2020A&A...633A.150H} {633, A150}

\bibitem[\protect\citeauthoryear{{Johnson}, {Rudnick}, {Jones}, {Mendygral}  \&
  {Dolag}}{{Johnson} et~al.}{2020}]{Johnson2020}
{Johnson} A.~R.,  {Rudnick} L.,  {Jones} T.~W.,  {Mendygral} P.~J.,   {Dolag}
  K.,  2020, \mndoi [\apj] {10.3847/1538-4357/ab5d30}, \href
  {https://ui.adsabs.harvard.edu/abs/2020ApJ...888..101J} {888, 101}

\bibitem[\protect\citeauthoryear{{Johnston-Hollitt} et~al.,}{{Johnston-Hollitt}
  et~al.}{2015}]{JH2015}
{Johnston-Hollitt} M.,  et~al., 2015, Advancing Astrophysics with the Square
  Kilometre Array (AASKA14), \href
  {http://adsabs.harvard.edu/abs/2015aska.confE..92J} {p.~92}

\bibitem[\protect\citeauthoryear{{Kharb}, {Shastri}  \& {Gabuzda}}{{Kharb}
  et~al.}{2005}]{Kharb2005}
{Kharb} P.,  {Shastri} P.,   {Gabuzda} D.~C.,  2005, \mndoi [\apjl]
  {10.1086/497984}, \href
  {https://ui.adsabs.harvard.edu/abs/2005ApJ...632L..69K} {632, L69}

\bibitem[\protect\citeauthoryear{{Knuettel}, {Gabuzda}  \&
  {O'Sullivan}}{{Knuettel} et~al.}{2017}]{Knuettel2017}
{Knuettel} S.,  {Gabuzda} D.,   {O'Sullivan} S.,  2017, \mndoi [Galaxies]
  {10.3390/galaxies5040061}, \href
  {https://ui.adsabs.harvard.edu/abs/2017Galax...5...61K} {5, 61}

\bibitem[\protect\citeauthoryear{{Kriek}, {van der Wel}, {van Dokkum}, {Franx}
  \& {Illingworth}}{{Kriek} et~al.}{2008}]{Kriek2008}
{Kriek} M.,  {van der Wel} A.,  {van Dokkum} P.~G.,  {Franx} M.,
  {Illingworth} G.~D.,  2008, \mndoi [\apj] {10.1086/589677}, \href
  {https://ui.adsabs.harvard.edu/abs/2008ApJ...682..896K} {682, 896}

\bibitem[\protect\citeauthoryear{{Kronberg} \& {Simard-Normandin}}{{Kronberg}
  \& {Simard-Normandin}}{1976}]{KS1976}
{Kronberg} P.~P.,  {Simard-Normandin} M.,  1976, \mndoi [\nat]
  {10.1038/263653a0}, \href
  {https://ui.adsabs.harvard.edu/abs/1976Natur.263..653K} {263, 653}

\bibitem[\protect\citeauthoryear{{Kuiper} et~al.,}{{Kuiper}
  et~al.}{2011}]{Kuiper2011}
{Kuiper} E.,  et~al., 2011, \mndoi [\mnras] {10.1111/j.1365-2966.2011.18852.x},
  \href {https://ui.adsabs.harvard.edu/abs/2011MNRAS.415.2245K} {415, 2245}

\bibitem[\protect\citeauthoryear{{Kurk}, {R{\"o}ttgering}, {Pentericci}  \&
  {Miley}}{{Kurk} et~al.}{2000}]{Kurk2000}
{Kurk} J.,  {R{\"o}ttgering} H.,  {Pentericci} L.,   {Miley} G.,  2000, in
  {Mazure} A.,  {Le F{\`e}vre} O.,   {Le Brun} V.,  eds,  Astronomical Society
  of the Pacific Conference Series Vol. 200, Clustering at High Redshift.
  p.~424 (\mn@eprint {arXiv} {astro-ph/9910257})

\bibitem[\protect\citeauthoryear{{Kurk}, {Pentericci}, {R{\"o}ttgering}  \&
  {Miley}}{{Kurk} et~al.}{2002}]{Kurk2002}
{Kurk} J.~D.,  {Pentericci} L.,  {R{\"o}ttgering} H.~J.~A.,   {Miley} G.~K.,
  2002, in {Henney} W.~J.,  {Steffen} W.,  {Binette} L.,   {Raga} A.,  eds,
  Revista Mexicana de Astronomia y Astrofisica Conference Series Vol. 13,
  Revista Mexicana de Astronomia y Astrofisica Conference Series. pp 191--195
  (\mn@eprint {arXiv} {astro-ph/0102337})

\bibitem[\protect\citeauthoryear{{Laing}}{{Laing}}{1980}]{Laing1980}
{Laing} R.~A.,  1980, \mndoi [\mnras] {10.1093/mnras/193.3.439}, \href
  {http://adsabs.harvard.edu/abs/1980MNRAS.193..439L} {193, 439}

\bibitem[\protect\citeauthoryear{{Lyutikov}, {Pariev}  \& {Gabuzda}}{{Lyutikov}
  et~al.}{2005}]{Lyutikov2005}
{Lyutikov} M.,  {Pariev} V.~I.,   {Gabuzda} D.~C.,  2005, \mndoi [\mnras]
  {10.1111/j.1365-2966.2005.08954.x}, \href
  {http://adsabs.harvard.edu/abs/2005MNRAS.360..869L} {360, 869}

\bibitem[\protect\citeauthoryear{{Macquart}, {Ekers}, {Feain}  \&
  {Johnston-Hollitt}}{{Macquart} et~al.}{2012}]{Macquart2012}
{Macquart} J.-P.,  {Ekers} R.~D.,  {Feain} I.,   {Johnston-Hollitt} M.,  2012,
  \mndoi [\apj] {10.1088/0004-637X/750/2/139}, \href
  {http://adsabs.harvard.edu/abs/2012ApJ...750..139M} {750, 139}

\bibitem[\protect\citeauthoryear{Madau \& Dickinson}{Madau \&
  Dickinson}{2014}]{MD2014}
Madau P.,  Dickinson M.,  2014, \mndoi [Annual Review of Astronomy and
  Astrophysics] {10.1146/annurev-astro-081811-125615}, 52, 415

\bibitem[\protect\citeauthoryear{{Mahmud}, {Coughlan}, {Murphy}, {Gabuzda}  \&
  {Hallahan}}{{Mahmud} et~al.}{2013}]{Mahmud2013}
{Mahmud} M.,  {Coughlan} C.~P.,  {Murphy} E.,  {Gabuzda} D.~C.,   {Hallahan}
  D.~R.,  2013, \mndoi [\mnras] {10.1093/mnras/stt201}, \href
  {http://adsabs.harvard.edu/abs/2013MNRAS.431..695M} {431, 695}

\bibitem[\protect\citeauthoryear{{Marsden}, {Shankar}, {Ginolfi}  \&
  {Zubovas}}{{Marsden} et~al.}{2020}]{Marsden2020}
{Marsden} C.,  {Shankar} F.,  {Ginolfi} M.,   {Zubovas} K.,  2020, \mndoi
  [Frontiers in Physics] {10.3389/fphy.2020.00061}, \href
  {https://ui.adsabs.harvard.edu/abs/2020FrP.....8...61M} {8, 61}

\bibitem[\protect\citeauthoryear{{McCarthy}}{{McCarthy}}{1993}]{McCarthy1993}
{McCarthy} P.~J.,  1993, \mndoi [\araa] {10.1146/annurev.aa.31.090193.003231},
  \href {https://ui.adsabs.harvard.edu/abs/1993ARA&A..31..639M} {31, 639}

\bibitem[\protect\citeauthoryear{{McCarthy}, {Spinrad}, {van Breugel},
  {Liebert}, {Dickinson}, {Djorgovski}  \& {Eisenhardt}}{{McCarthy}
  et~al.}{1990}]{McCarthy1990}
{McCarthy} P.~J.,  {Spinrad} H.,  {van Breugel} W.,  {Liebert} J.,  {Dickinson}
  M.,  {Djorgovski} S.,   {Eisenhardt} P.,  1990, \mndoi [\apj]
  {10.1086/169503}, \href
  {https://ui.adsabs.harvard.edu/abs/1990ApJ...365..487M} {365, 487}

\bibitem[\protect\citeauthoryear{{Miley} \& {De Breuck}}{{Miley} \& {De
  Breuck}}{2008}]{MD2008}
{Miley} G.,  {De Breuck} C.,  2008, \mndoi [\aapr] {10.1007/s00159-007-0008-z},
  \href {https://ui.adsabs.harvard.edu/abs/2008A&ARv..15...67M} {15, 67}

\bibitem[\protect\citeauthoryear{{Miley} et~al.,}{{Miley}
  et~al.}{2006}]{Miley2006}
{Miley} G.~K.,  et~al., 2006, \mndoi [\apjl] {10.1086/508534}, \href
  {https://ui.adsabs.harvard.edu/abs/2006ApJ...650L..29M} {650, L29}

\bibitem[\protect\citeauthoryear{{Mukherjee}, {Bicknell}, {Wagner},
  {Sutherland}  \& {Silk}}{{Mukherjee} et~al.}{2018}]{Mukherjee2018}
{Mukherjee} D.,  {Bicknell} G.~V.,  {Wagner} A.~Y.,  {Sutherland} R.~S.,
  {Silk} J.,  2018, \mndoi [\mnras] {10.1093/mnras/sty1776}, \href
  {https://ui.adsabs.harvard.edu/abs/2018MNRAS.479.5544M} {479, 5544}

\bibitem[\protect\citeauthoryear{Muldrew, Hatch  \& Cooke}{Muldrew
  et~al.}{2015}]{Muldrew2015}
Muldrew S.~I.,  Hatch N.~A.,   Cooke E.~A.,  2015, \mndoi [Monthly Notices of
  the Royal Astronomical Society] {10.1093/mnras/stv1449}, 452, 2528

\bibitem[\protect\citeauthoryear{{Murgia}, {Govoni}, {Feretti}, {Giovannini},
  {Dallacasa}, {Fanti}, {Taylor}  \& {Dolag}}{{Murgia}
  et~al.}{2004}]{Murgia2004}
{Murgia} M.,  {Govoni} F.,  {Feretti} L.,  {Giovannini} G.,  {Dallacasa} D.,
  {Fanti} R.,  {Taylor} G.~B.,   {Dolag} K.,  2004, \mndoi [\aap]
  {10.1051/0004-6361:20040191}, \href
  {https://ui.adsabs.harvard.edu/abs/2004A&A...424..429M} {424, 429}

\bibitem[\protect\citeauthoryear{{Murphy} \& {Gabuzda}}{{Murphy} \&
  {Gabuzda}}{2012}]{Murphy2012}
{Murphy} E.,  {Gabuzda} D.~C.,  2012, in Journal of Physics Conference Series.
  p. 012009 (\mn@eprint {arXiv} {1109.4778}),
  \mndoi{10.1088/1742-6596/355/1/012009}

\bibitem[\protect\citeauthoryear{{Murphy} et~al.,}{{Murphy}
  et~al.}{2018}]{Murphy2018}
{Murphy} E.~J.,  et~al., 2018, in {Murphy} E.,  ed.,  Astronomical Society of
  the Pacific Conference Series Vol. 517, Science with a Next Generation Very
  Large Array. p.~3 (\mn@eprint {arXiv} {1810.07524})

\bibitem[\protect\citeauthoryear{{Narayan}, {Chael}, {Chatterjee}, {Ricarte}
  \& {Curd}}{{Narayan} et~al.}{2021}]{Narayan2021}
{Narayan} R.,  {Chael} A.,  {Chatterjee} K.,  {Ricarte} A.,   {Curd} B.,  2021,
  arXiv e-prints, \href {https://ui.adsabs.harvard.edu/abs/2021arXiv210812380N}
  {p. arXiv:2108.12380}

\bibitem[\protect\citeauthoryear{{Neronov} \& {Vovk}}{{Neronov} \&
  {Vovk}}{2010}]{NV2010}
{Neronov} A.,  {Vovk} I.,  2010, \mndoi [Science] {10.1126/science.1184192},
  \href {https://ui.adsabs.harvard.edu/abs/2010Sci...328...73N} {328, 73}

\bibitem[\protect\citeauthoryear{{Nesvadba}, {De Breuck}, {Lehnert}, {Best},
  {Binette}  \& {Proga}}{{Nesvadba} et~al.}{2011}]{Nesvadba2011}
{Nesvadba} N.~P.~H.,  {De Breuck} C.,  {Lehnert} M.~D.,  {Best} P.~N.,
  {Binette} L.,   {Proga} D.,  2011, \mndoi [\aap]
  {10.1051/0004-6361/201014960}, \href
  {https://ui.adsabs.harvard.edu/abs/2011A&A...525A..43N} {525, A43}

\bibitem[\protect\citeauthoryear{{Nesvadba}, {Drouart}, {De Breuck}, {Best},
  {Seymour}  \& {Vernet}}{{Nesvadba} et~al.}{2017}]{Nesvadba2017}
{Nesvadba} N.~P.~H.,  {Drouart} G.,  {De Breuck} C.,  {Best} P.,  {Seymour} N.,
    {Vernet} J.,  2017, \mndoi [\aap] {10.1051/0004-6361/201629357}, \href
  {https://ui.adsabs.harvard.edu/abs/2017A&A...600A.121N} {600, A121}

\bibitem[\protect\citeauthoryear{{O'Sullivan} et~al.,}{{O'Sullivan}
  et~al.}{2012}]{OSullivan2012}
{O'Sullivan} S.~P.,  et~al., 2012, \mndoi [\mnras]
  {10.1111/j.1365-2966.2012.20554.x}, \href
  {http://adsabs.harvard.edu/abs/2012MNRAS.421.3300O} {421, 3300}

\bibitem[\protect\citeauthoryear{{O'Sullivan} et~al.,}{{O'Sullivan}
  et~al.}{2013}]{OSullivan2013b}
{O'Sullivan} S.~P.,  et~al., 2013, \mndoi [\apj] {10.1088/0004-637X/764/2/162},
  \href {http://adsabs.harvard.edu/abs/2013ApJ...764..162O} {764, 162}

\bibitem[\protect\citeauthoryear{{O'Sullivan}, {Combes}, {Babul}, {Chapman},
  {Phadke}, {Schellenberger}  \& {Salom{\'e}}}{{O'Sullivan}
  et~al.}{2021}]{OSullivan2021}
{O'Sullivan} E.,  {Combes} F.,  {Babul} A.,  {Chapman} S.,  {Phadke} K.~A.,
  {Schellenberger} G.,   {Salom{\'e}} P.,  2021, \mndoi [\mnras]
  {10.1093/mnras/stab2825}, \href
  {https://ui.adsabs.harvard.edu/abs/2021MNRAS.508.3796O} {508, 3796}

\bibitem[\protect\citeauthoryear{Offringa, McKinley, Hurley-Walker
  et~al.}{Offringa et~al.}{2014}]{Offringa2014}
Offringa A.~R.,  McKinley B.,  Hurley-Walker  et~al., 2014, \mndoi [MNRAS]
  {10.1093/mnras/stu1368}, 444, 606

\bibitem[\protect\citeauthoryear{{Paolillo}, {Fabbiano}, {Peres}  \&
  {Kim}}{{Paolillo} et~al.}{2002}]{Paolillo2002}
{Paolillo} M.,  {Fabbiano} G.,  {Peres} G.,   {Kim} D.-W.,  2002, \mndoi [\apj]
  {10.1086/337919}, \href {http://adsabs.harvard.edu/abs/2002ApJ...565..883P}
  {565, 883}

\bibitem[\protect\citeauthoryear{{Pasetto}, {Carrasco-Gonz{\'a}lez},
  {O'Sullivan}, {Basu}, {Bruni}, {Kraus}, {Curiel}  \& {Mack}}{{Pasetto}
  et~al.}{2018}]{Pasetto2018}
{Pasetto} A.,  {Carrasco-Gonz{\'a}lez} C.,  {O'Sullivan} S.,  {Basu} A.,
  {Bruni} G.,  {Kraus} A.,  {Curiel} S.,   {Mack} K.-H.,  2018, preprint, \href
  {http://adsabs.harvard.edu/abs/2018arXiv180109731P} {} (\mn@eprint {arXiv}
  {1801.09731})

\bibitem[\protect\citeauthoryear{{Pentericci}, {Roettgering}, {Miley},
  {Carilli}  \& {McCarthy}}{{Pentericci} et~al.}{1997}]{Pentericci1997}
{Pentericci} L.,  {Roettgering} H.~J.~A.,  {Miley} G.~K.,  {Carilli} C.~L.,
  {McCarthy} P.,  1997, \aap, \href
  {https://ui.adsabs.harvard.edu/abs/1997A&A...326..580P} {326, 580}

\bibitem[\protect\citeauthoryear{{Pentericci}, {R{\"o}ttgering}, {Miley},
  {McCarthy}, {Spinrad}, {van Breugel}  \& {Macchetto}}{{Pentericci}
  et~al.}{1999}]{Pentericci99}
{Pentericci} L.,  {R{\"o}ttgering} H.~J.~A.,  {Miley} G.~K.,  {McCarthy} P.,
  {Spinrad} H.,  {van Breugel} W.~J.~M.,   {Macchetto} F.,  1999, \aap, \href
  {https://ui.adsabs.harvard.edu/abs/1999A&A...341..329P} {341, 329}

\bibitem[\protect\citeauthoryear{Perley \& Butler}{Perley \&
  Butler}{2013}]{Perley2013}
Perley R.~A.,  Butler B.~J.,  2013, \mndoi [The Astrophysical Journal
  Supplement Series] {10.1088/0067-0049/206/2/16}, 206, 16

\bibitem[\protect\citeauthoryear{{Planck Collaboration} et~al.,}{{Planck
  Collaboration} et~al.}{2020}]{Planck2020}
{Planck Collaboration} et~al., 2020, \mndoi [\aap]
  {10.1051/0004-6361/201833910}, \href
  {https://ui.adsabs.harvard.edu/abs/2020A&A...641A...6P} {641, A6}

\bibitem[\protect\citeauthoryear{{Reuland} et~al.,}{{Reuland}
  et~al.}{2007}]{Reuland2007}
{Reuland} M.,  et~al., 2007, \mndoi [\aj] {10.1086/516571}, \href
  {http://adsabs.harvard.edu/abs/2007AJ....133.2607R} {133, 2607}

\bibitem[\protect\citeauthoryear{{Russell} et~al.,}{{Russell}
  et~al.}{2017}]{Russell2017}
{Russell} H.~R.,  et~al., 2017, \mndoi [\apj] {10.3847/1538-4357/836/1/130},
  \href {http://adsabs.harvard.edu/abs/2017ApJ...836..130R} {836, 130}

\bibitem[\protect\citeauthoryear{{Sault}, {Teuben}  \& {Wright}}{{Sault}
  et~al.}{1995}]{Sault1995}
{Sault} R.~J.,  {Teuben} P.~J.,   {Wright} M.~C.~H.,  1995, in {Shaw} R.~A.,
  {Payne} H.~E.,   {Hayes} J.~J.~E.,  eds,  Astronomical Society of the Pacific
  Conference Series Vol. 77, Astronomical Data Analysis Software and Systems
  IV. p.~433 (\mn@eprint {} {astro-ph/0612759})

\bibitem[\protect\citeauthoryear{{Sault}, {Hamaker}  \& {Bregman}}{{Sault}
  et~al.}{1996}]{Sault1996}
{Sault} R.~J.,  {Hamaker} J.~P.,   {Bregman} J.~D.,  1996, \aaps, \href
  {https://ui.adsabs.harvard.edu/abs/1996A&AS..117..149S} {117, 149}

\bibitem[\protect\citeauthoryear{{Taylor} \& {Perley}}{{Taylor} \&
  {Perley}}{1993}]{TP1993}
{Taylor} G.~B.,  {Perley} R.~A.,  1993, \mndoi [\apj] {10.1086/173257}, \href
  {http://adsabs.harvard.edu/abs/1993ApJ...416..554T} {416, 554}

\bibitem[\protect\citeauthoryear{{Tozzi} et~al.,}{{Tozzi}
  et~al.}{2022}]{Tozzi2022}
{Tozzi} P.,  et~al., 2022, arXiv e-prints, \href
  {https://ui.adsabs.harvard.edu/abs/2022arXiv220302208T} {p. arXiv:2203.02208}

\bibitem[\protect\citeauthoryear{{Vazza} et~al.,}{{Vazza}
  et~al.}{2021a}]{Vazza2021}
{Vazza} F.,  et~al., 2021a, \mndoi [Galaxies] {10.3390/galaxies9040109}, \href
  {https://ui.adsabs.harvard.edu/abs/2021Galax...9..109V} {9, 109}

\bibitem[\protect\citeauthoryear{{Vazza}, {Wittor}, {Brunetti}  \&
  {Br{\"u}ggen}}{{Vazza} et~al.}{2021b}]{Vazza_2021}
{Vazza} F.,  {Wittor} D.,  {Brunetti} G.,   {Br{\"u}ggen} M.,  2021b, \mndoi
  [\aap] {10.1051/0004-6361/202140513}, \href
  {https://ui.adsabs.harvard.edu/abs/2021A&A...653A..23V} {653, A23}

\bibitem[\protect\citeauthoryear{{Villar-Mart{\'\i}n}, {Vernet}, {di Serego
  Alighieri}, {Fosbury}, {Humphrey}  \& {Pentericci}}{{Villar-Mart{\'\i}n}
  et~al.}{2003}]{VM2003}
{Villar-Mart{\'\i}n} M.,  {Vernet} J.,  {di Serego Alighieri} S.,  {Fosbury}
  R.,  {Humphrey} A.,   {Pentericci} L.,  2003, \mndoi [\mnras]
  {10.1046/j.1365-2966.2003.07090.x}, \href
  {https://ui.adsabs.harvard.edu/abs/2003MNRAS.346..273V} {346, 273}

\bibitem[\protect\citeauthoryear{{Vogt} \& {En{\ss}lin}}{{Vogt} \&
  {En{\ss}lin}}{2005}]{VE2005}
{Vogt} C.,  {En{\ss}lin} T.~A.,  2005, \mndoi [\aap]
  {10.1051/0004-6361:20041839}, \href
  {http://adsabs.harvard.edu/abs/2005A%26A...434...67V} {434, 67}

\bibitem[\protect\citeauthoryear{{Walter} et~al.,}{{Walter}
  et~al.}{2020}]{Fabian2020}
{Walter} F.,  et~al., 2020, \mndoi [\apj] {10.3847/1538-4357/abb82e}, \href
  {https://ui.adsabs.harvard.edu/abs/2020ApJ...902..111W} {902, 111}

\bibitem[\protect\citeauthoryear{{Weinberger} et~al.,}{{Weinberger}
  et~al.}{2017a}]{Weinberger2017b}
{Weinberger} R.,  et~al., 2017a, \mndoi [\mnras] {10.1093/mnras/stw2944}, \href
  {http://adsabs.harvard.edu/abs/2017MNRAS.465.3291W} {465, 3291}

\bibitem[\protect\citeauthoryear{{Weinberger}, {Ehlert}, {Pfrommer}, {Pakmor}
  \& {Springel}}{{Weinberger} et~al.}{2017b}]{Weinberger2017}
{Weinberger} R.,  {Ehlert} K.,  {Pfrommer} C.,  {Pakmor} R.,   {Springel} V.,
  2017b, \mndoi [\mnras] {10.1093/mnras/stx1409}, \href
  {http://adsabs.harvard.edu/abs/2017MNRAS.470.4530W} {470, 4530}

\bibitem[\protect\citeauthoryear{Wu, Ghisellini, Hodges-Kluck, Gallo, Ciardi,
  Haardt, Sbarrato  \& Tavecchio}{Wu et~al.}{2017}]{Wu2017}
Wu J.,  Ghisellini G.,  Hodges-Kluck E.,  Gallo E.,  Ciardi B.,  Haardt F.,
  Sbarrato T.,   Tavecchio F.,  2017, \mndoi [Monthly Notices of the Royal
  Astronomical Society] {10.1093/mnras/stx416}, 468, 109

\bibitem[\protect\citeauthoryear{{Zhuravleva}, {Allen}, {Mantz}  \&
  {Werner}}{{Zhuravleva} et~al.}{2018}]{Zhuravleva2018}
{Zhuravleva} I.,  {Allen} S.~W.,  {Mantz} A.,   {Werner} N.,  2018, \mndoi
  [\apj] {10.3847/1538-4357/aadae3}, \href
  {https://ui.adsabs.harvard.edu/abs/2018ApJ...865...53Z} {865, 53}

\bibitem[\protect\citeauthoryear{{Zirm} et~al.,}{{Zirm}
  et~al.}{2005}]{Zirm2005}
{Zirm} A.~W.,  et~al., 2005, \mndoi [\apj] {10.1086/431921}, \href
  {https://ui.adsabs.harvard.edu/abs/2005ApJ...630...68Z} {630, 68}

\bibitem[\protect\citeauthoryear{{van Ojik}}{{van Ojik}}{1995}]{vO1995}
{van Ojik} R.,  1995, PhD thesis, University of Leiden, Netherlands

\makeatother
\end{thebibliography}

\end{document}